\documentclass{aa}
\usepackage{txfonts,latexsym,graphicx,natbib,amssymb}

\usepackage{color}
\usepackage{url}

\newcommand\given[1][]{\:#1\vert\:}
\newcommand\sigintr{\sigma_\mathrm{intr}}
\newcommand\muintr{\mu_\mathrm{intr}}
\newcommand\FeHobs{\mathrm{[Fe/H]}_\mathrm{obs}}
\newcommand\FeHobsi{\mathrm{[Fe/H]}_{\mathrm{obs},i}}
\newcommand\FeH{\mathrm{[Fe/H]}}

\newcommand\likelihood{\mathcal{L}}
\newcommand\param{\boldsymbol{\theta}}
\newcommand{\Sgren}{Str{\"o}mgren~}

\begin{document}

\title{\Sgren $uvby$ photometry of the peculiar globular cluster NGC~2419\thanks{Based on observations made with the Isaac Newton Telescope operated on the island of La Palma by the Isaac Newton Group in the Spanish Observatorio del Roque de los Muchachos of the Instituto de Astrofísica de Canarias. The photometric catalogue is available in electronic form
at the CDS via anonymous ftp to cdsarc.u-strasbg.fr or via http://cdsweb.u-strasbg.fr/cgi-bin/qcat?J/A+A/}}
      
\author{ Matthias J. Frank\inst{1}\thanks{E-mail: mfrank@lsw.uni-heidelberg.de}
        \and Andreas Koch\inst{1} 
        \and Sofia Feltzing\inst{2} 
        \and Nikolay Kacharov\inst{1}\thanks{Present address: European Southern Observatory, Karl-Schwarzschild-Straße 2, 85748, Garching bei München, Germany}
        \and Mark I. Wilkinson\inst{3} 
        \and Mike Irwin\inst{4}
}
\institute{
        Landessternwarte, Zentrum f\"ur Astronomie der Universit\"at Heidelberg, K\"onigsstuhl 12, D-69117 Heidelberg, Germany
        \and Lund Observatory, Box 43, SE-22100 Lund, Sweden
        \and Department of Physics and Astronomy, University of Leicester, University Road, Leicester LE1 7RH, United Kingdom
        \and Institute of Astronomy, Madingley Road, Cambridge CB03 0HA, United Kingdom
}

\date{Received ?????; accepted ?????}

\abstract{NGC~2419 is a peculiar Galactic globular cluster offset from the others in the size-luminosity diagram, and showing several chemical abundance anomalies. Here, we present \Sgren $uvby$ photometry of the cluster. Using the gravity- and metallicity-sensitive $c_1$ and $m_1$ indices, we identify a sample of likely cluster members extending well beyond the formal tidal radius. The estimated contamination by cluster non-members is only one per cent, making our catalogue ideally suited for spectroscopic follow-up.
We derive photometric [Fe/H] of red giants, and depending on which metallicity calibration from the literature we use, we find reasonable to excellent agreement with spectroscopic [Fe/H], both for the cluster mean metallicity and for individual stars. We demonstrate explicitly that the photometric uncertainties are not Gaussian and this must be accounted for in any analysis of the metallicity distribution function. Using a realistic, non-Gaussian model for the photometric uncertainties, we find a formal internal [Fe/H] spread of $\sigma=0.11^{+0.02}_{-0.01}$\,dex. This is an upper limit to the cluster's true [Fe/H] spread and may partially, and possibly entirely, reflect the limited precision of the photometric metallicity estimation and systematic effects. The lack of correlation between spectroscopic and photometric [Fe/H] of individual stars is further evidence against a [Fe/H] spread on the 0.1\,dex level. 
Finally, the CN-sensitive $\delta_4$, among other colour indices, anti-correlates strongly with magnesium abundance, indicating that the second-generation stars are nitrogen enriched. The absence of similar correlations in some other CN-sensitive indices supports the second generation being enriched in He, which in these indices approximately compensates the shift due to CN. Compared to a single continuous distribution with finite dispersion, the observed $\delta_4$ distribution of red giants is slightly better fit by two distinct populations with no internal spread, with the nitrogen-enhanced second generation accounting for $53\pm5$\,per cent of stars. 
Despite its known peculiarities, NGC~2419 appears to be very similar to other metal-poor Galactic globular clusters with a similarly nitrogen-enhanced second generation and little or no variation in [Fe/H], which sets it apart from other suspected accreted nuclei such as $\omega$Cen.
}


\maketitle

\section{Introduction}
\label{sec:intro}
NGC~2419, with a Galactocentric distance of $95\pm3$\,kpc\footnote{Throughout this paper we adopt a distance modulus of 19.71$\pm$0.08\,mag for NGC~2419, measured by \citet{2011AJ....141...81D} based on RR Lyrae; the distance of the Sun to the Galactic centre is assumed to be 8.3\,kpc \citep{2009ApJ...692.1075G}.} is one of the Milky Way's outermost globular clusters (GCs), and is highly unusual among the Galactic GC population in terms of structure and chemistry. Its absolute magnitude of M$_V=-9.4$\,mag \citep[][2010 edition]{1996AJ....112.1487H}, places it among the brightest Galactic GCs and at the same time, it is one of the most extended clusters, with $r_h=24\pm1$\,pc \citep[][]{2007A&A...473..171B}, compared to the average Galactic GC with r$_h\sim5$\,pc. 
Thus, it has been suggested that NGC~2419 is the remnant core of an accreted former Milky Way satellite \citep[e.g.][]{2004MNRAS.354..713V}, although to date no firm link with Galactic halo substructure, such as the Virgo, Gemini, Orphan, or Sagittarius streams has been established \citep{2009ApJ...701L..29C,2013ApJ...765..154D,2011ApJ...729...69B,2014ApJ...784...19C,2014MNRAS.437..116B}.
Given its unique location in the size-luminosity plane, it has also been suggested that NGC~2419 is a local example at the low-surface-brightness end of the ultra-compact dwarf galaxy population \citep{2011AJ....142..199B}. 
However,  the nature of these objects  remains ambiguous, as this morphological class likely constitutes a mixture of stripped nuclei and of massive genuine star clusters \citep[see][for a recent review]{2014AN....335..486F}. 

In contrast to $\omega$\,Cen (NGC~5139) or M54, which are thought to be remnant nuclei of accreted satellites and which have broad metallicity distributions ($\omega$Cen: [Fe/H]=$-1.75$ to $-0.75$\,dex, \citealt{2010ApJ...722.1373J}; M54: intrinsic r.m.s. scatter in [Fe/H]$\sim0.19$\,dex, \citealt{2010A&A...520A..95C}), the case for a spread in the abundance of heavy elements in NGC~2419 is less clear. Several recent studies based on medium- to high-resolution spectra of a few to a few tens of red giant branch (RGB) stars in NGC~2419 found no spread in [Fe/H] above what is expected from measurement errors, but suggested a possible spread of $\sim\!0.2$\,dex in the calcium abundance \citep[][]{2011ApJ...740...60C,2012ApJ...760...86C,2012MNRAS.426.2889M}. A spread in [Ca/H] is also supported by the results of \citet{2013ApJ...778L..13L}, who showed that NGC~2419's bright RGB stars show a spread, and a hint of a bimodality, in the $hk$ photometric index (constructed from \Sgren $Ca$, $b$, and $y$ magnitudes, the first of which is centred on the Ca II K and H lines). Interestingly, these authors also find that a small spread in metallicity of [Fe/H]$\sim\!0.2$\,dex best explains the observations.

Regarding light elements, observations of NGC~2419 revealed a more exotic chemical inventory. Its horizontal branch (HB) morphology suggests a subpopulation (accounting for $\sim$\!30\,per cent of the stars) that is strongly enriched in He \citep{2008AJ....136.2259S,2011MNRAS.414.3381D}. 
Studying the distribution of HB stars in ultraviolet and optical \textit{Hubble Space Telescope} photometry, \citet{2015MNRAS.446.1469D} found evidence of three populations of HB stars differing in Helium content, the `regular' blue HB with solar He (Y=0.25), an extreme population (EHB) with Y$\sim$0.36, and a smaller population of intermediate HB stars that can be reproduced with an intermediate He content (Y$\sim$0.28). Similarly, an intrinsic broadening of the RGB in broadband colours was also interpreted as being due to a spread in He of $\Delta \mathrm{Y}\sim0.1$ \citep{2011MNRAS.414.3381D,2013MNRAS.431.1995B,2013ApJ...778L..13L}.

Most strikingly, two chemically distinct populations of bright RGB stars in NGC~2419 have been found \citep{2011ApJ...740...60C,2012ApJ...760...86C,2012MNRAS.426.2889M}. One population is moderately enhanced in Mg ([Mg/Fe]$\,\sim\!$+0.5\,dex), has roughly solar K, and is similar to red giants in other GCs and in the Galactic halo. The other population, about 40\% in the sample of \citet{2012MNRAS.426.2889M}, is strongly depleted in Mg (down to [Mg/Fe]$\sim$-1\,dex) and enhanced in K ([K/Fe]$\,\sim\,$+1.5\,dex). These two populations produce a pronounced Mg-K anti-correlation in NGC~2419. In this respect NGC~2419 appears unique among Galactic GCs \citep{2013ApJ...769...40C}, even though \citet{2015ApJ...801...68M} recently found a similar, but less extreme population of Mg-deficient, K-enriched stars in NGC~2808, which might indicate that a Mg-K anti-correlation is a common pattern in massive GCs. 

In the framework of multiple stellar populations in GCs, light-element variations are understood to be the result of self-enrichment in GCs, where one or more second generations of stars form from, or accrete, material polluted by the first generation\footnote{We follow the terminology that refers to the enriched GC stars as `second generation', and use the terms `populations' and `generations' interchangeably, but we note that not all of the proposed enrichment scenarios actually invoke distinct star formation episodes or an age difference like the term `generation' may suggest \citep[e.g. the disc accretion scenario of][]{2013MNRAS.436.2398B}.} of stars (see e.g. \citealt{2012A&ARv..20...50G} and \citealt{2015MNRAS.449.3333B}, for an overview and critical discussion of different scenarios). In addition to the well-known Na-O anti-correlation, where the second-generation stars are Na-rich and O-poor, enrichment in Al and N and depletion in Mg and C are also common in second-generation GC stars. 

In  this paper we present \Sgren $uvby$ photometry of NGC~2419. Originally devised for the study of early-type main-sequence stars, the \Sgren photometric system \citep{1963QJRAS...4....8S,1964ApNr....9..333S} was subsequently extended to later spectral types, lower metallicities, and red giants \citep[e.g.][]{1989A&A...221...65S,1994AJ....107.1577A}. Following early studies of the nearest GCs \citep[][]{1983A&A...128..194A,1987AJ.....93.1454A}, \Sgren imaging has been obtained for a large number of Galactic GCs. These studies, for example,  focused  on calibrating relations between the \Sgren metallicity index $m_1$ and spectroscopic [Fe/H] \citep[e.g.][]{2000A&A...355..994H,2007ApJ...670..400C}, on metallicity and age distributions in $\omega$Cen \citep[][]{2000AJ....119.1225H,2000A&A...362..895H,2009ApJ...706.1277C}, and distance-independent age-estimates \citep[][]{1998ApJ...500L.179G,2000AJ....120.1884G,2002A&A...395..481G}. 

Furthermore \Sgren $uvby$ photometry is sensitive to variations in nitrogen and carbon \citep[][]{1980ApJS...44..517B,1994AJ....107.1577A,2000AJ....120.3111A} owing to the 3400\AA\ NH and the 4316\AA\ CN molecular absorption bands that lie in the $u$ and $v$ band-passes, respectively. Consequently, \Sgren photometry has received recent attention in the context of the aforementioned multiple stellar generations in GCs \citep[e.g.][]{2008ApJ...684.1159Y,2011A&A...534A...9S,2011A&A...535A.121C,2013MmSAI..84...63A}.

This paper is structured as follows. Our observations, basic data reduction, and photometry are described in Section~\ref{sec:obs}. Section ~\ref{sec:membership} discusses the selection and classification of cluster member stars, which is then used in Section~\ref{sec:photmet} to measure photometric metallicities on the RGB and to study the cluster's intrinsic metallicity distribution function. In Section~\ref{sec:multiplepop} we study light-element patterns in NGC~2419. Our results are summarised and discussed in Section~\ref{sec:disc}.

\section{Observations}
\label{sec:obs}

\begin{table}
\caption{Log of observations.}
\label{tab:observations}     
\centering  
\begin{tabular}{c c c c c}         
\hline\hline                       
Filter &  Exp. time & Seeing$^{a}$ &  Airmass &     MJD \\
& [min] & [arcsec] & & \\
\hline
$y$ &             45 &         1.9 &1.06 & 55975.89 \\
$u$ &             45 &         1.9 &1.02 & 55975.93 \\
$b$ &             29 &         2.6 &1.11 & 55976.03 \\
$b$ &             45 &         3.4 &1.48 & 55976.10 \\
$v$ &             45 &         2.4 &1.12 & 55977.85 \\
$y$ &             45 &         1.6 &1.02 & 55977.91 \\
$b$ &             45 &         2.0 &1.02 & 55977.96 \\
$y$ &             45 &         1.9 &1.10 & 55978.01 \\
$v$ &              7 &         2.5 &1.18 & 55978.05 \\
$b$ &             45 &         2.2 &1.29 & 55978.06 \\
$b$ &             45 &         2.5 &1.50 & 55978.09 \\
\hline                                            
\end{tabular}

$^{a}$Median measured FWHM of isolated stars on the images.
\end{table}

\begin{table}
\caption{Observed standard stars with magnitudes and color indices from \citet{1988A&AS...73..225S}, except for star 521110099, whose measurements are from \citet{1994A&AS..106..257O}.}
\label{tab:standardstars}
\centering  
\begin{tabular}{r r r r r}          
\hline\hline                        
 uvby98 ID$^{a}$ &       V &    $b-y$ &     $m_1$ &     $c_1$ \\
  &       [mag] &    [mag] &    [mag] &     [mag]\\
\hline
   -102457 &   9.765 &  0.606 &  0.555 &  0.152 \\
   -403208 &   9.998 &  0.311 &  0.048 &  0.373 \\
   -503063 &   9.734 &  0.568 &  0.461 &  0.182 \\
   -503763 &  10.239 &  0.579 &  0.546 &  0.241 \\
  -1303834 &  10.685 &  0.415 &  0.098 &  0.183 \\
   5101696 &   9.912 &  0.397 &  0.100 &  0.180 \\
 980014039 &  12.828 &  0.587 &  0.267 &  0.153 \\
   2602606 &   9.731 &  0.336 &  0.050 &  0.280 \\
 100100363 &   8.648 &  0.191 &  0.139 &  0.760 \\
 521110099 &   9.081 &  0.317 &  0.157 &  0.483 \\
 100108754 &   9.006 &  0.435 &  0.217 &  0.254 \\
 100118659 &   8.827 &  0.422 &  0.196 &  0.244 \\
 100123265 &   8.348 &  0.504 &  0.356 &  0.348 \\
 100134088 &   7.992 &  0.392 &  0.137 &  0.255 \\
 100051754 &   9.000 &  0.375 &  0.144 &  0.290 \\
 100064090 &   8.279 &  0.428 &  0.110 &  0.126 \\
 100088371 &   8.414 &  0.407 &  0.186 &  0.329 \\
\hline                                            
\end{tabular}

$^a$Homogenized identifier according to \citet[][]{1998A&AS..129..431H}.
\end{table}

\begin{figure}
\includegraphics[width=\linewidth]{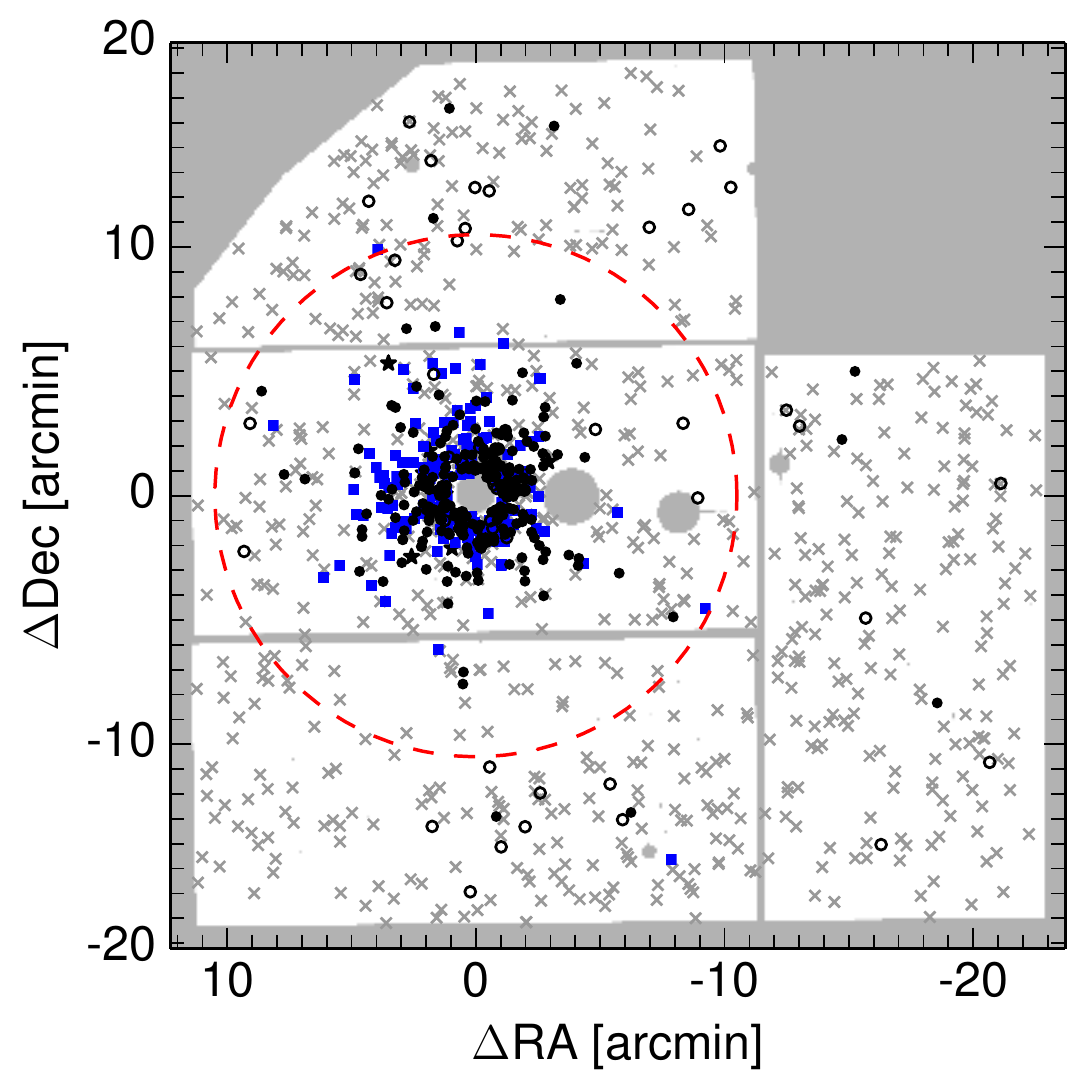}
\caption{Sources on the sky (north is up, east is left) and their classification (see Section~\ref{sec:membership}). Filled blue squares indicate candidate HB members; filled dots likely RGB/AGB members; open dots RGB/AGB-star candidates, based on the $c_0$ selection, that were rejected by  a metallicity criterion; and grey crosses  likely cluster non-members. Grey areas are masked out and reflect chip gaps, vignetting, and areas heavily contaminated by bright stars. The dashed red circle has a radius of 10.5\,arcmin (267\,pc at the distance of NGC~2419), beyond which we assume detections to be non-member stars for the purpose of estimating the contamination in our cluster member sample in Section~\ref{sec:purity}.}
\label{fig:onsky}
\end{figure}

Our imaging was obtained on two nights in February 2012 using the Wide Field Camera (WFC) at the 2.5m Isaac Newton Telescope (INT) at the Roque de los Muchachos Observatory, La Palma, Spain. 
The WFC's wide field of view ($34\arcmin\times34\arcmin$) allowed us to trace the large extent of NGC~2419 out to $\sim$4 times its formal limiting radius \citep[$r_t=7.1\pm1.0\,\arcmin$;][]{2007A&A...473..171B}, and out to $2.5$ times the maximum radius of 10.5\,\arcmin at which member stars have been reported \citep{2007ApJ...667L..61R}. 
Images of NGC~2419 were taken in the intermediate-band \Sgren filters $u$, $b$, $v$, and $y$ and the cluster was placed near the centre of the camera's central chip 4 (Fig.\ref{fig:onsky}). Our individual science exposures are listed in Table~\ref{tab:observations}. In between the science observations, we observed standard stars selected from the catalogues of \citet{1988A&AS...73..225S} and \citet{1994A&AS..106..257O}. These are listed in Table~\ref{tab:standardstars}.

The basic data reduction was carried out using the INT Wide Field Survey pipeline \citep{2001NewAR..45..105I} operated at the Cambridge Astronomical Survey Unit. 
The pipeline handles bias removal and  flat-fielding (using twilight flats obtained during the run), corrects for non-linearity and gain differences, and provides an astrometric solution. 

The individual images were dithered by small offsets (5-12\,arcsec) to mitigate detector cosmetics. 
Since the offsets were not large enough to bridge the chip gaps ($\sim\!17$\,arcsec), we obtained photometry of the camera's four chips independently. Instrumental magnitudes were obtained via PSF-fitting using the \textsc{Daophot} software packages 
\citep{1987PASP...99..191S}. 
The PSF was constructed from 80-150 non-saturated, bright, relatively isolated stars on each frame and we allowed for a quadratic variation in the PSF across the frame. 
The pipeline-provided astrometry was used as a first guess on the geometric transformation between individual frames, which was then refined using \textsc{Daomaster} \citep{1993spct.conf..291S}. 
A detection image was created using \textsc{Daophot}'s \textsc{mosaic2} to stack the three $y$-band images that had the best image quality (see Table~\ref{tab:observations}).

Simultaneous PSF-fitting photometry of all science frames of a given chip was then performed using \textsc{Allframe} \citep{1994PASP..106..250S}.
Aperture photometry in a series of apertures was obtained for the standard stars and for  the PSF stars on the science frames (with neighbouring stars subtracted), and \textsc{Daogrow} \citep{1990PASP..102..932S} was used to derive aperture corrections. 
We verified that the aperture corrections on a given frame did not significantly vary with position on the frame, suggesting that the quadratically varying model PSF captures the spatial variation of the actual PSF.

The resulting instrumental magnitudes were transformed to the \Sgren system using calibration equations of the form 
\begin{equation}
\label{eq:photcal}
y_\mathrm{std} = y_\mathrm{obs} + \mathrm{ZP}_{y,\mathrm{night}} + \beta_{y,\mathrm{night}}\,(X-1)+\gamma_y\,(v-y)_\mathrm{std}, 
\end{equation} and analogously for $u$, $v$, and $b$, using $(v-y)$ for the colour term in all transformations. Here $X$ is the effective airmass. The zero points $\mathrm{ZP}_{x,\mathrm{night}}$ and extinction coefficients $\beta_{x,\mathrm{night}}$ were derived for both nights independently. The colour term coefficient $\gamma_{x}$ was fit for both nights simultaneously, assuming that the colour-correction onto the standard system is primarily a function of the filter transmission and camera sensitivity rather than of atmospheric conditions. 
 
Upon merging the individual frame catalogues, we found small systematic differences between the calibrated magnitudes of the stars in common on several $v$, $b$, or $y$ frames (in $u$ there is only one science exposure). We therefore applied a per frame correction of the zero point in order to bring all frames in a given filter to a common (median) flux scale. The r.m.s. scatter of these median magnitude corrections between frames was $\sim$\!0.03 mag, which gives an estimate of the precision of our zeropoints and thereby of the accuracy of our photometry relative to the standard system. While both nights were reported as largely photometric by the Carlsberg Meridian Telescope on La Palma\footnote{http://www.ast.cam.ac.uk/$\sim\!$dwe/SRF/camc\_extinction.html}, the presence of these small offsets indicate that the two nights were not perfectly photometric throughout, although no consistent time-dependent trend could be seen in the residuals of Eq.~\ref{eq:photcal} and its $uvb$ analogues. 

We note that by using the standard star catalogue by \citet{1988A&AS...73..225S} our photometry is linked to the system of secondary \Sgren standards published by \citet{1993A&AS..102...89O}. A second major system of secondary \Sgren standards for metal-poor giants exists and is anchored to the catalogues by \citet{1980ApJS...44..517B} and \citet{1994AJ....107.1577A}; see \citet{2007A&A...465..357F} for a discussion and \citet{1995A&A...295..710O} for a comparison between the two systems.

In order to retain only reliably measured, bona fide stars in our catalogue, we examined plots of the \textsc{Daophot} $\mathrm{sharpness}$ and $\chi$ diagnostics versus magnitude, and required the following criteria to hold: $\chi<2$, and a statistical uncertainty of $\sigma<0.2$\,mag in all four filters; $\lvert\mathrm{sharpness}\rvert<1$ in $u$, $v$, and $b$, and $\lvert\mathrm{sharpness}\rvert<0.5$ in $y$ where the seeing was best, allowing for a better removal of extended sources. 
This produced a catalogue with 1192 sources in the entire WFC field of view with measurements in all four bands. The catalogue does not cover the central region of NGC~2419 (r$\la0.85$\,arcmin), where our imaging is strongly affected by crowding. The photometric catalogue, including the classification of sources (Section~\ref{sec:membership}), is available at CDS. 

In order to de-redden the measured magnitudes, we adopted a reddening of $E(B-V)=0.08\pm0.01$\,mag towards NGC~2419 \citep{2011AJ....141...81D}. For the bandpass specific extinction coefficients $A_x/\mathrm{E(B-V)}$ we adopted the values for $R_V=3.1$ from \citet{2011ApJ...737..103S}: 4.305, 3.793, 3.350, and 2.686 for $x=u$, $v$, $b$, and $y$, respectively. In the following, we exclusively use de-reddened magnitudes and colours.

\begin{figure*}
\label{fig:c0vsbmy}
\centering
\includegraphics[width=0.8\linewidth]{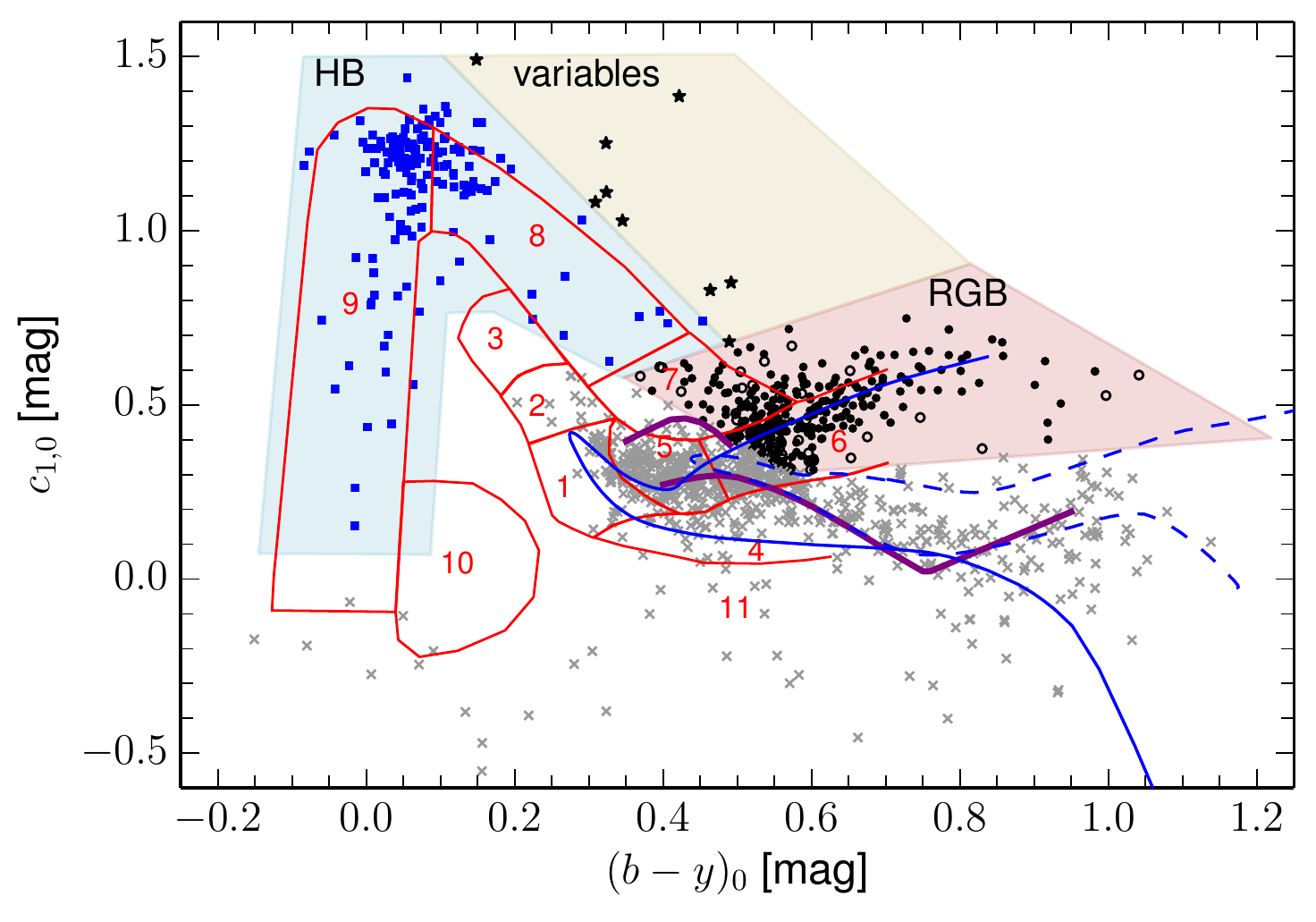}
\caption{Surface-gravity-sensitive $c_{1,0}$ index vs. $(b-y)_0$ of objects in our catalogue. Different symbols represent likely cluster RGB/AGB stars (filled circles), candidate RGB/AGB stars that do not fulfil the metallicity selection (open circles; see Fig.~\ref{fig:m0vsbmy}), likely cluster HB stars (filled blue squares), likely cluster variable stars (filled star symbols), and foreground contaminants as well as background galaxies (grey crosses; see text). Shaded regions indicate the selection boxes for RGB (red), HB (blue), and variable stars (yellow). Superimposed are the empirical dwarf star sequence for [Fe/H]=0\,dex (lower purple line) and the upper envelope of dwarf stars in the solar neighbourhood (upper purple line) derived by \citet{2010A&A...521A..40A}, as well as 13\,Gyr Dartmouth isochrones for [Fe/H]=0\,dex (blue dashed line) and [Fe/H]=-2\,dex. Thin red lines delineate the loci of different stages in the evolution of metal-poor stars and were adopted from fig. 6 of Schuster et al. (2004). The regions correspond to: 1 turn-off stars, 2 blue straggler / turn-off transition, 3 blue stragglers, 4 main sequence, 5 subgiant stars, 6 RGB stars, 7 red HB / AGB transition, 8 HB, 9 blue HB, 10 subluminous / blue HB transition, 11 subluminous stars.}
\end{figure*}

Extensive artificial star tests were performed to quantify the photometric uncertainties in our data. For these, we generated lists of stars distributed on the sky in a regularly spaced grid (with step size 13\,arcsec in order to avoid self-crowding in the artificial star tests); the origin of the grid was moved for each artificial star experiment by a random offset, leading - on average - to uniform coverage on the sky. The colours and magnitudes of the bulk of the artificial stars were chosen to be similar to the observed stars. This was done by smoothing the observed distributions in $u-y$, $v-y$, and $b-y$ colours and $y$-band magnitude with a Gaussian of 0.2\,mag dispersion and drawing artificial stars from this distribution. To map the completeness and saturation limits of our photometry, a small fraction ($\sim\!10$\,per cent) of artificial stars was drawn from uniform colour- and magnitude distributions extending well beyond those of the observed stars.

The resulting input catalogues were then transformed to instrumental magnitudes and positions on each science frame by inverting the photometric calibration and geometric transformations, and the stars were inserted into the science frames using the empirical PSF. Photometry was repeated on these sets of frames in exactly the same way as for the original science frames. The resulting photometry was cross-matched with the superset of the original photometric catalogue and the input artificial star catalogue. Artificial stars were counted as recovered if they were detected within 3 pixels (approximately half of the PSF FWHM in the best-seeing frames) of the inserted position. The final artificial star catalogue contained $\sim2.6\times10^5$ simulated  stars, of which $\sim1.4\times10^5$ were recovered and passed the same quality cuts in $\chi$ and $\mathrm{sharpness}$ as the observed stars. Photometric uncertainties as a function of magnitudes and position on the sky were then derived from the differences between inserted and recovered magnitudes (see Section~\ref{sec:photmet}).

\section{Cluster membership}
\label{sec:membership}
\Sgren photometry provides an excellent way to discriminate between foreground and background contamination and our target cluster's RGB, HB, and asymptotic giant branch (AGB) stars \citep[see][]{2010A&A...521A..40A}. Recent applications to the Draco and Hercules dwarf spheroidals can be found in \citet{2007A&A...465..357F} and \citet{2009A&A...506.1147A}, respectively. 

\subsection{Candidate RGB, AGB, HB, and variable star cluster members}
Our selection of cluster members relies primarily on the gravity-sensitive \Sgren $c_{1} = (u-v) - (v-b)$ index that is shown as a function of $(b-y)_0$ colour in Fig.~\ref{fig:c0vsbmy}. We classified sources in our catalogue into likely HB stars, candidate RGB and AGB stars, and likely variable stars, using selection boxes shown as shaded regions in this diagram. 

The design of the selection boxes was guided by the following empirical and theoretical sequences that are also shown in Fig.~\ref{fig:c0vsbmy}: Red lines and numbers show the empirical loci occupied by metal-poor stars at different evolutionary stages from the main-sequence to the AGB \citep[adopted from][]{2004A&A...422..527S}. Also shown are the empirical sequence of dwarf stars for solar metallicity (lower purple line), and the upper envelope of the regime of dwarf stars in the solar neighbourhood (upper purple line), derived by \citet{2010A&A...521A..40A}. The two isochrones represent old stellar populations (13\,Gyr, solar [$\alpha$/Fe]) of solar metallicity (dashed blue line) and [Fe/H]=-2\,dex (solid blue line), respectively. These were taken from the Dartmouth Stellar Evolution Library \citet{2008ApJS..178...89D}, where we specifically used the 2008 version of the \Sgren isochrone tables that are based on the semi-empirical colour-temperature relations of \citet{2004AJ....127.1227C}.

The HB and bright RGB of NGC~2419 are immediately obvious in Fig.~\ref{fig:c0vsbmy}, as is the sequence of foreground dwarfs. It is also apparent that the discrimination of HB stars against the foreground is relatively unambiguous, while for the RGB, contamination by dwarfs and subgiants for sources bluer than $(b-y)_0\la0.6$\,mag cannot be avoided. Moreover, there is no obvious distinction in the distribution of our observed sources between RGB, AGB, and red HB stars. 

Stars lying significantly above the RGB/AGB and HB sequences are classified as candidate variables \citep[see][]{AdenThesis}. In fact, five out of the nine sources in the variables selection box can be identified with confirmed variable stars published by \citet{2011AJ....141...81D}.  

There are several sources that fall into the subluminous region, below the well-defined sequences ($c_{1,0}\la0$\,mag). Among the redder of these sources, starting at $(b-y)_0\ga0.3$\,mag, several are clearly visible as background galaxies on the best-seeing $y$-band images. Since there will also be more compact background objects in the field, this indicates that a good fraction of the detections below the dwarf sequence in Fig.~\ref{fig:c0vsbmy} may be background galaxies. The bluer subluminous detections with $(b-y)_0\la0.3$\,mag and $c_0\la0$\,mag, appear to be point sources that are not concentrated around NGC~2419 on the sky. Some of these sources may be white dwarfs that are known to occupy this region in the $c_1 - (b-y)$ diagram \citep[e.g.][]{1983AJ.....88..109W}.

\subsection{Likely RGB and AGB star cluster members}
\label{sec:likelyrgb}
\begin{figure}
\includegraphics[width=\linewidth]{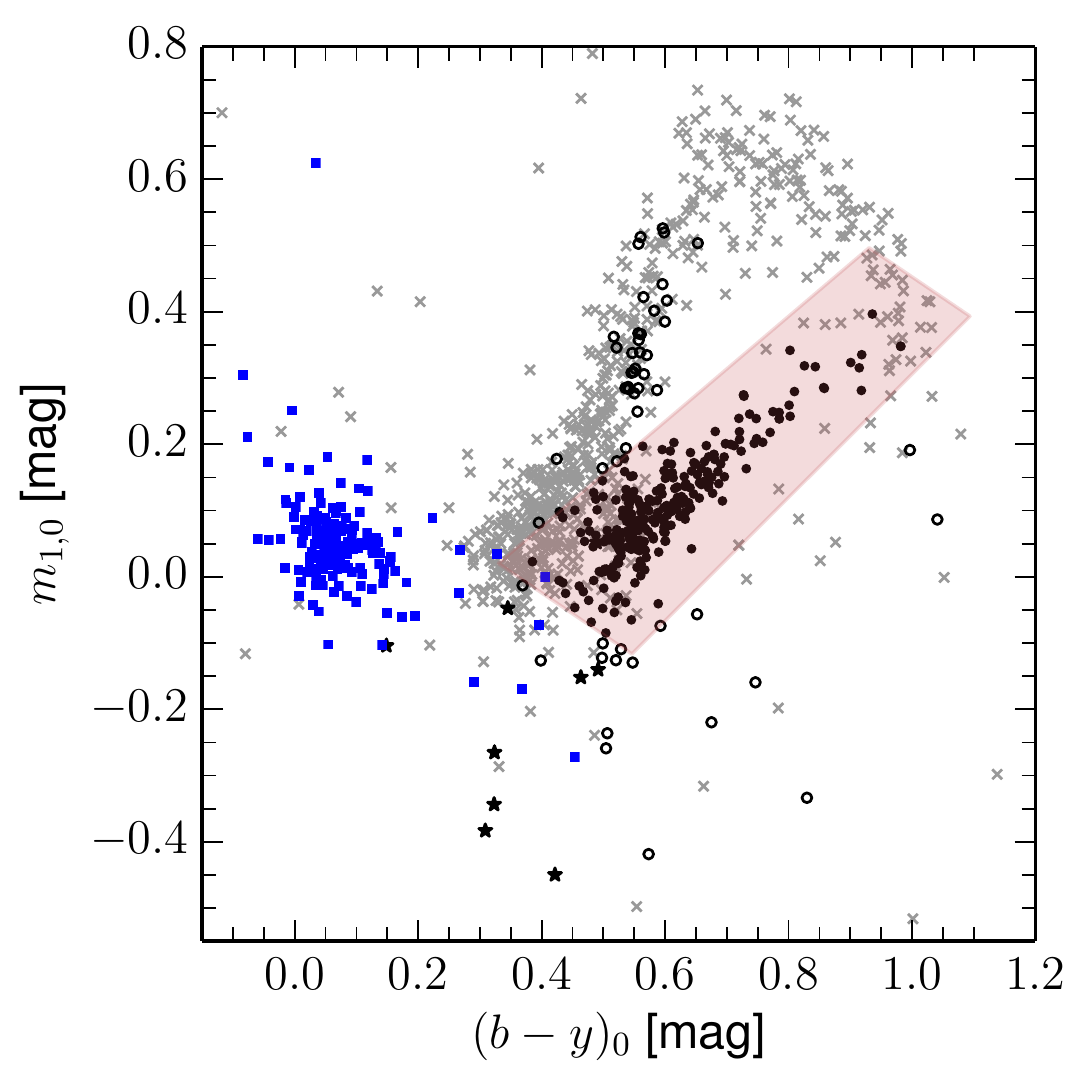}
\caption{Metallicity index $m_{1,0}$ index vs. $(b-y)_0$. Symbols are as in Fig.~\ref{fig:c0vsbmy}. The shaded red area represents the crude selection in metallicity that was used to suppress the contamination in the range $(b-y)_0<0.6$\,mag, where dwarfs and giants are not clearly separated in the $c_{1,0}$ vs. $(b-y)_0$ diagram (see Fig.~\ref{fig:c0vsbmy}).}
\label{fig:m0vsbmy}
\end{figure}

In order to suppress contamination in the RGB/AGB sample, a crude selection based on the metallicity index $m_1=(v-b)-(b-y) $ index was applied. Figure~\ref{fig:m0vsbmy} shows the $m_{1,0}$ vs. $(b-y)_0$ plane, where the curves of constant metallicity are approximately straight lines (see Section~\ref{sec:photmet}). Our selection box in this diagram serves to exclude higher metallicity contaminants from the candidate RGB/AGB sample bluewards of $(b-y)_0\la0.6$\,mag. 

\begin{figure}
\includegraphics[width=\linewidth]{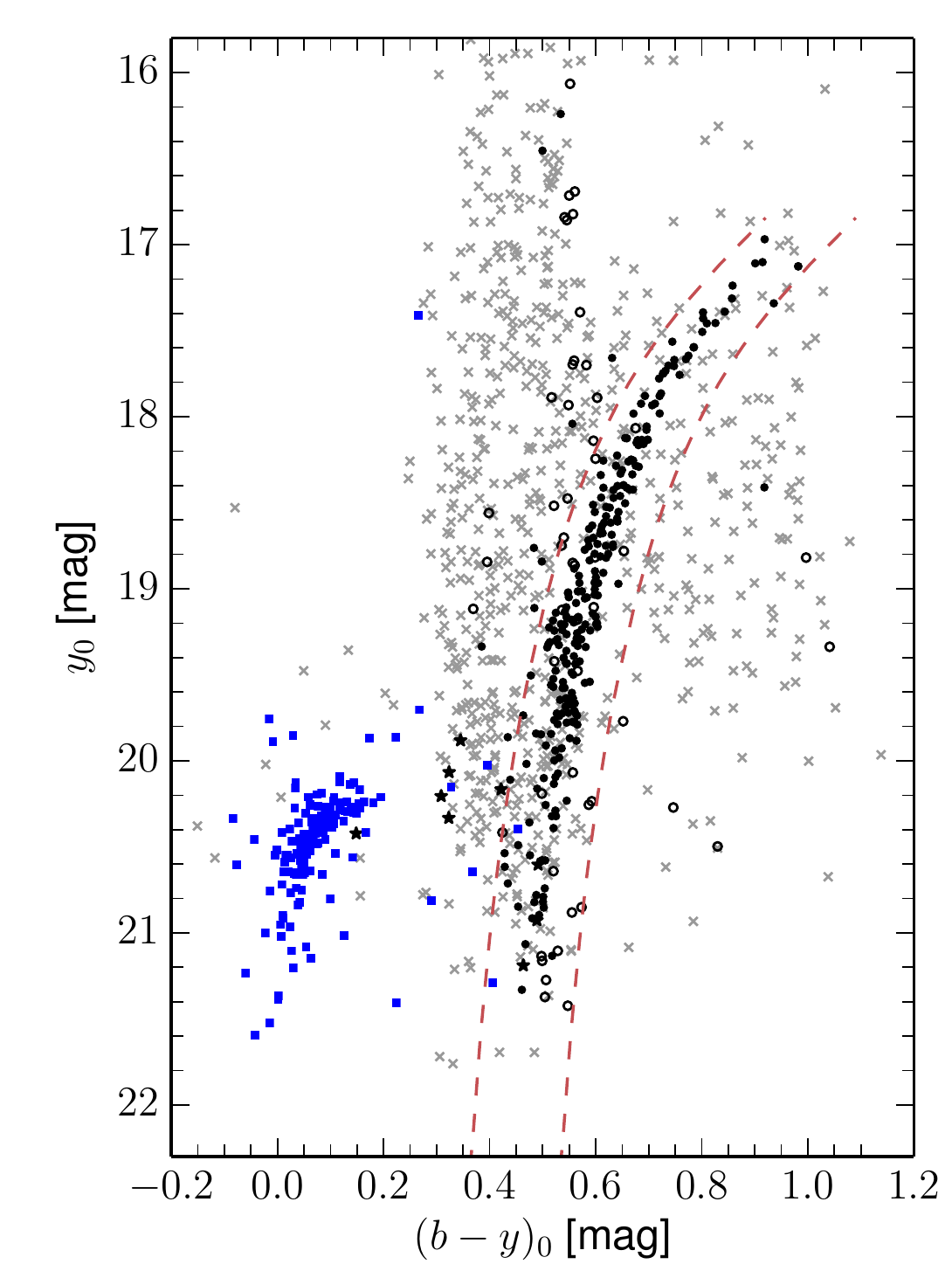}
\caption{$y_0$ vs. $(b-y)_0$ colour-magnitude diagram with symbols as in Fig.~\ref{fig:c0vsbmy}. It is apparent that the selection of candidate cluster members in the $c_1$ and $m_1$ vs. $(b-y)$ colour-colour diagrams (Figs.~\ref{fig:c0vsbmy} and ~\ref{fig:m0vsbmy}) is very efficient. To obtain a clean sample of RGB stars we use an additional CMD-based selection indicated by the red dashed lines. These were obtained by shifting a 13\,Gyr, [Fe/H]=-2\,dex, [$\alpha$/Fe]=0.4\,dex Dartmouth isochrone from its best-fitting location by -0.04 and 0.13\,mag along the colour axis in order to eliminate several AGB star candidates, as well as a handful of foreground and background contaminants.}
\label{fig:cmdyvsbmy}
\end{figure}

\subsection{A clean sample of RGB stars}
\label{sec:cleanrgb}
The selection in the $c_{1,0}$ and $m_{1,0}$ vs. $(b-y)_0$ colour planes is independent of distance. To incorporate distance information, we used an additional  selection based  on ridge lines in the $y_0$ vs. $(b-y)_0$ colour-magnitude diagram, shown in Fig.~\ref{fig:cmdyvsbmy}. This yields a clean sample of 251 RGB stars (with a few possible AGB contaminants) that will be used to study the metallicity distribution and multiple populations in NGC~2419 in Sects.~\ref{sec:photmet} and \ref{sec:multiplepop}. Figure~\ref{fig:cmdyvsbmy} also demonstrates the excellent cleaning that is already achieved using only the two colour-colour diagrams. Examining the positions on the sky of the obvious outliers in this diagram, we find only four sources far away from the cluster centre: the two brightest RGB/AGB candidates ($y_0\sim$16.3\,mag), both probably foreground turn-off stars based on their location in the $c_{1,0}$ vs. $(b-y)_0$ diagram; the RGB/AGB candidate at $y_0\sim18.4$\,mag, $(b-y)_0\sim0.9$\,mag, a potential background galaxy, or evolved AGB halo star \citep[given its similar location to the carbon star in Fig.~9 of][]{2007A&A...465..357F}; and the HB-candidate with $y_0\sim$17.3\,mag, a likely foreground HB star based on its location in Fig.~\ref{fig:c0vsbmy}.

\subsection{Purity of the selection}
\label{sec:purity}
The distribution of sources in our catalogue on the sky is shown in Fig.~\ref{fig:onsky}. In this figure, grey regions represent the areas falling off the WFC's four chips, the vignetted corner of chip 3 (north-east), as well as areas where completeness is heavily affected by bright stars in the field or crowding in the centre of NGC~2419. While sources classified as likely non-members (grey crosses) are dispersed across the field, our likely RGB/AGB stars (filled dots), HB stars (filled blue squares), and variable stars (star symbols) are concentrated  around the cluster centre. 

A quantitative estimate of the purity of the cluster member samples can be made by assuming that all stars beyond a cluster-centric radius of $r=10.5$\,arcmin (illustrated by the red dashed circle in Fig.~\ref{fig:onsky}) are non-members. This is the maximum radius at which the deep ground-based CMDs published by \citet{2007ApJ...667L..61R} clearly show the upper main-sequence and turn-off of NGC~2419. The effective area (excluding the grey regions in Fig.~\ref{fig:onsky}) covered by our photometry is 330\,arcmin$^2$ inside r=10.5\,arcmin and 640\,arcmin$^2$ beyond that radius. 

Out  of the entire candidate cluster member sample, i.e. likely RGB/AGB, HB, and variable stars, 10 sources are located in the outer region and 419 sources in the inner region. Scaling by the relative areas covered, we thus expect $\sim5$ contaminants in our candidate cluster member sample, corresponding to a contamination fraction of 1.2\,per cent. Similarly, for the HB  and the clean RGB (i.e. RGB/AGB stars falling within the CMD-based selection in Fig.~\ref{fig:cmdyvsbmy}) samples alone, the estimated contamination is 0.6 and 0.8\,per cent, respectively.

We note that our sample is less complete than it is pure. Apart from photometric incompleteness, which could be quantified using the artificial star tests, there is no direct way of estimating the incompleteness of e.g. our `clean RGB' selection. Qualitatively, there is an overdensity of sources classified as non-members at the cluster's position on the sky, most of which lie on the fainter part of the RGB in Fig.~\ref{fig:cmdyvsbmy}. This suggests that a more complete sample, at the cost of a higher contamination by foreground dwarfs, could be obtained by using a more generous selection in the $(b-y)_0<0.6\,$mag region of Fig.~\ref{fig:c0vsbmy}. We refrain from doing so, since we are primarily interested in having a clean RGB star sample in the following analysis. The incompleteness due to this conservative selection of cluster giants with $(b-y)_0<0.6\,$mag will not bias our quantitative results on the metallicity spread and the chemical subpopulations, since we will restrict these analyses to stars brighter than $y_0$=19\,mag.

\section{Photometric metallicity}
\label{sec:photmet}
\begin{figure*}
\includegraphics[width=0.33\linewidth]{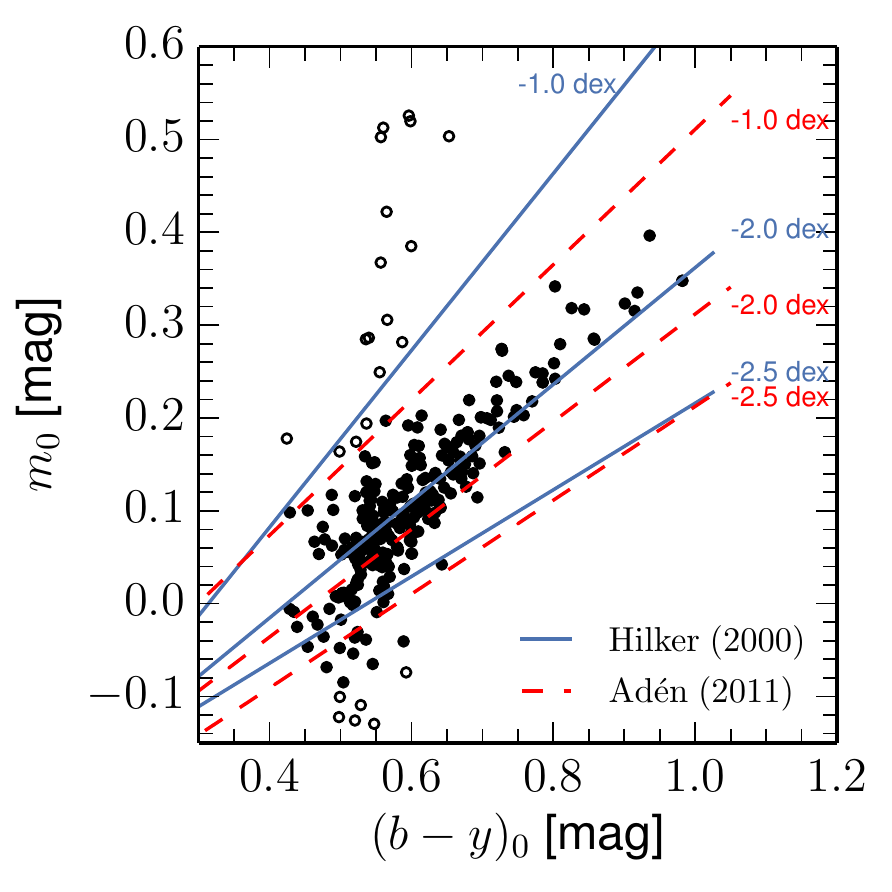}
\includegraphics[width=0.33\linewidth]{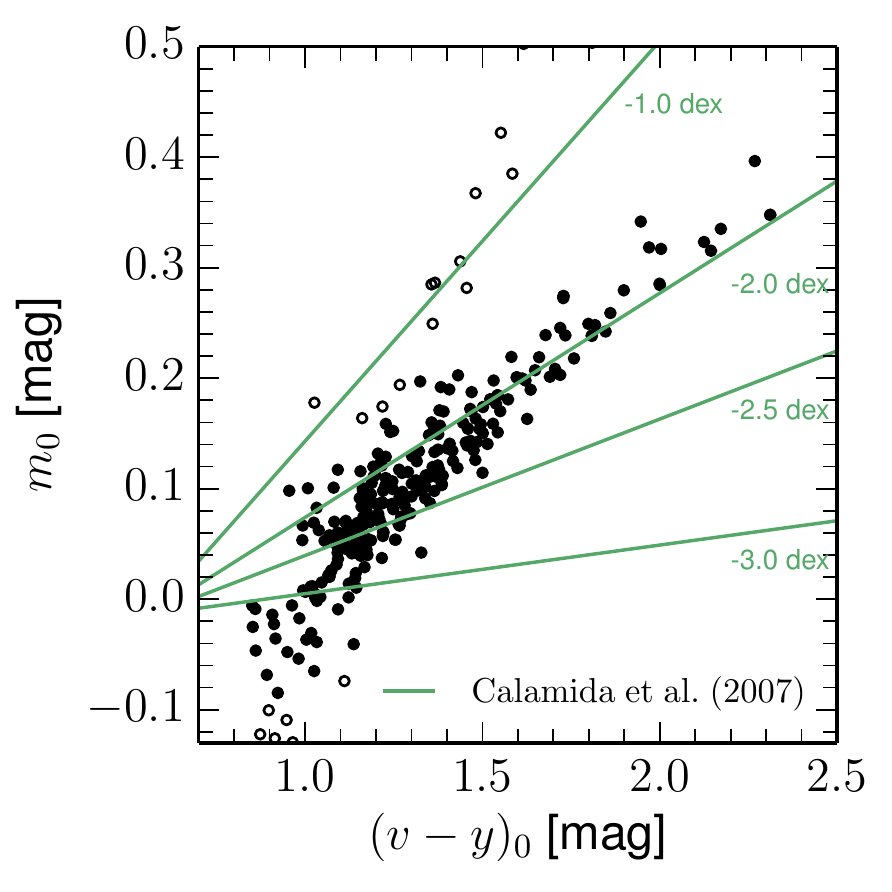}
\includegraphics[width=0.33\linewidth]{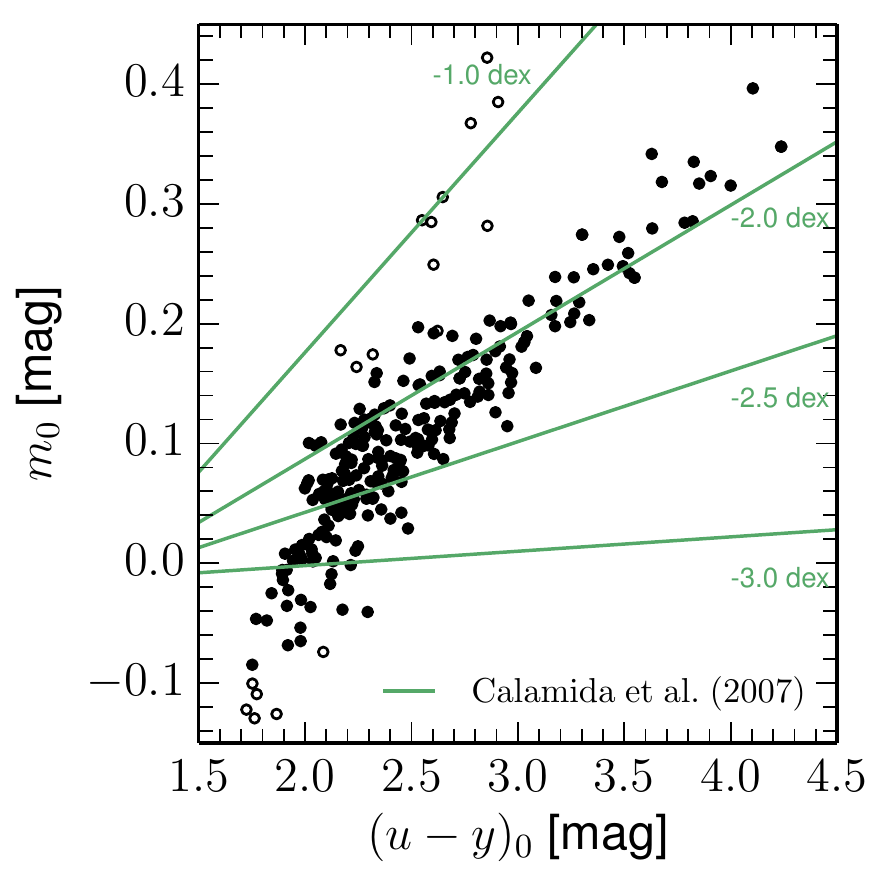}
\caption{Metallicity index $m_{1,0}$  vs. colour and lines of constant [Fe/H] according to metallicity calibrations from the literature. Filled dots represent the likely RGB star sample; open circles are the candidate RGB stars that were rejected based on the metallicity selection in Fig.~\ref{fig:m0vsbmy}, and serve to illustrate that this metallicity cut does not significantly bias our NGC~2419 sample. The left panel shows the calibration by \citet{2000A&A...355..994H} and \citet{AdenThesis} as a function of $(b-y)_0$ colour. The middle right panel shows the semi-empirical calibrations by \citet{2007ApJ...670..400C} as a function of $(v-y)_0$ and $(u-y)_0$, respectively.}
\label{fig:metallicityrelations}
\end{figure*}

Determining metallicities of RGB stars is a popular application of wide-field imaging in the \Sgren filters because of its high efficiency in terms of telescope time compared to spectroscopic observations. Therefore, numerous efforts have been made to calibrate the \Sgren $m_1$ index as a function of colour such as ($u$-$y$), ($v$-$y$), or ($b$-$y$) to spectroscopic metallicity measurements \citep[e.g.][]{1980ApJS...44..517B,1989A&A...211..199R,2000A&A...355..994H,2007ApJ...670..400C,2009ApJ...706.1277C}.

Here, we use the calibration relations from three different studies, all of which were designed to be valid down to low metallicities of [Fe/H]$\la-2$\,dex. Based on \Sgren photometry and spectroscopic metallicities of red giants in $\omega$Cen, M55, M22, and in the field, \citet{2000A&A...355..994H} derived an empirical metallicity relation for $m_1$ as a function of $(b-y)$ colour that is given by
\begin{equation}
\label{eq:photmetHilker}
\mathrm{[Fe/H]}_\mathrm{H00} = \frac{m_{1,0}+a_1(b-y)_0+a_2}{a_3(b-y)_0+a_4},
\end{equation}
with $a_1=-1.277\pm0.050$, $a_2=0.331\pm0.035$, $a_3=0.324\pm0.035$, and $a_4=-0.032\pm0.025$.

\citet{2007ApJ...670..400C} derived empirical metallicity relations for $m_1$ as a function of $(v-y)$ and $(u-y)$ colour, based on calibrators in five Galactic GCs with [Fe/H]=-2 to -0.7\,dex, as well as semi-empirical relations of the same form based on the theoretical stellar evolutionary tracks by \citet{2006ApJ...642..797P}, transformed to the \Sgren system using the empirical colour-temperature-relations by \citet{2004AJ....127.1227C}. We adopt the semi-empirical relations, since these are recommended by \citet{2007ApJ...670..400C} as being the most robust for estimating the metallicity of red giants. They are given by 
\begin{equation}
\label{eq:photmetCalamida07vy}
\mathrm{[Fe/H]}_\mathrm{C07vy} = \frac{m_{1,0}+a_1(v-y)_0+a_2}{a_3(v-y)_0+a_4},
\end{equation}
with coefficients $a_1=-0.521\pm0.001$, $a_2=0.309$, $a_3=0.159\pm0.001$, and $a_4=-0.09\pm0.002$, and 
\begin{equation}
\label{eq:photmetCalamida07uy}
\mathrm{[Fe/H]}_\mathrm{C07uy} = \frac{m_{1,0}+a_1(u-y)_0+a_2}{a_3(u-y)_0+a_4},
\end{equation} with coefficients $a_1=-0.294\pm0.002$, $a_2=0.323$, $a_3=0.094\pm0.001$, and $a_4=-0.099\pm0.005$.

Finally, \citet{AdenThesis} assembled spectroscopic metallicities of red giants in the Milky Way and in the dwarf spheroidals Draco, Sextans, Ursa Major II, and Hercules and combined them with \Sgren photometry to calibrate a relation for $m_1$ as a function of $(b-y)$ for low metallicities ($-3.3<$[Fe/H]$\la-1$\,dex) that is given by 
\begin{equation}
\label{eq:photmetAdenThesis}
\mathrm{[Fe/H]}_\mathrm{A11} = \frac{m_{1,0}+a_1(b-y)_0+a_2}{a_3(b-y)_0+a_4},
\end{equation} with coefficients $a_1=-0.878$, $a_2=0.168$, $a_3=0.147$, and $a_4=0.005$.

Figure~\ref{fig:metallicityrelations} shows the metallicity index $m_{1,0}$ vs. $(b-y)_0$, $(v-y)_0$, and $(u-y)_0$ of NGC~2419 RGB stars, together with lines of constant metallicity according the four calibrations (Eqs.~\ref{eq:photmetHilker} to ~\ref{eq:photmetAdenThesis}). We observe that the sequence of likely cluster red giants (filled dots) falls close to the [Fe/H]$=-2$\,dex curve of all four metallicity relations. While there is considerable scatter among the relations derived by different authors, the two relations by \citet{2007ApJ...670..400C} appear to be more consistent with each other. We also note  that the RGB appears slightly tilted with respect to all four metallicity relations. This likely results from a combination of the uncertainties in our photometric calibration, which we noted in Section~\ref{sec:obs}, and of the uncertainties in the metallicity relations. Finally, the intrinsic sensitivity to metallicity of all calibrations deteriorates at bluer colours. 

\begin{figure*}
\includegraphics[width=\linewidth]{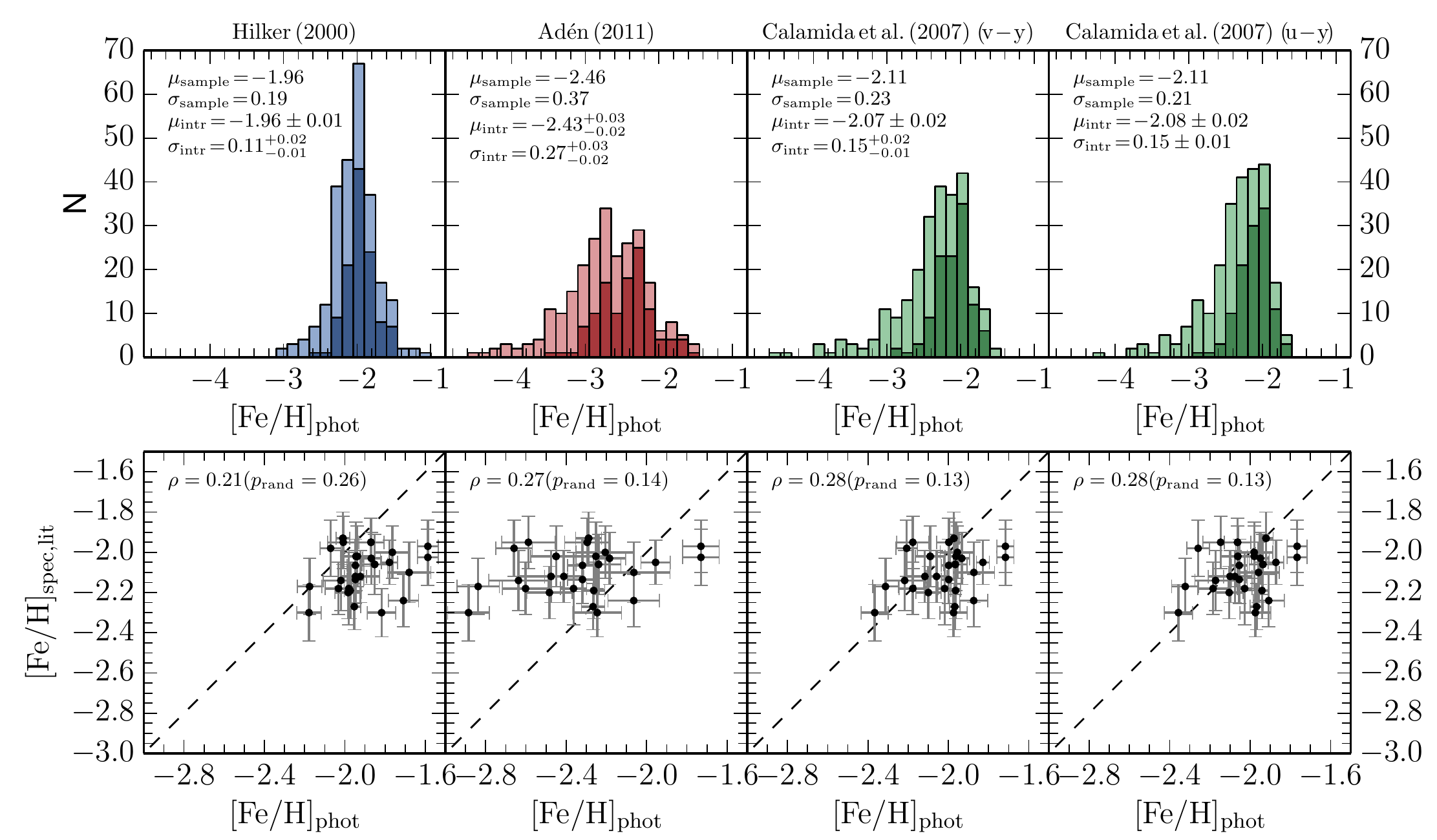}
\caption{Photometric metallicity distributions (upper panels) and comparison with spectroscopic measurements (lower panels). From left to right, the photometric metallicities were obtained using Eq.~\ref{eq:photmetHilker}, Eq.~\ref{eq:photmetAdenThesis}, Eq.~\ref{eq:photmetCalamida07vy}, and Eq.~\ref{eq:photmetCalamida07uy}. Light histograms in the upper panels show the entire clean RGB star sample, dark histograms show only stars with $y_0<19$\,mag. Each panel in the upper row reports, for the brighter subsample ($y_0<19$\,mag), the sample mean $\mu_\mathrm{sample}$ and standard deviation $\sigma_\mathrm{sample}$, as well as the intrinsic mean $\mu_\mathrm{intr}$ and dispersion $\sigma_\mathrm{intr}$ obtained via Eq.~\ref{eq:posteriorpdf}.}
\label{fig:metallicityhistograms}
\end{figure*}

Using Eq.~\ref{eq:photmetHilker} to ~\ref{eq:photmetAdenThesis}, we calculate the photometric [Fe/H] for all stars in our clean RGB sample (Section~\ref{sec:cleanrgb}). The resulting distributions are shown as the light histograms in the upper panels of Fig.~\ref{fig:metallicityhistograms}. Dark histograms show the metallicity distributions obtained using only stars brighter than $y_0=19$\,mag. This removes most of the metal-poor tail that is caused by lower metallicity sensitivity at bluer colours in combination with larger photometric uncertainties for fainter stars and by the systematic tilt of the observed RGB with respect to the curves of constant metallicity. Moreover, this excludes the magnitude range, where our metallicity-based pre-selection of member stars (Section~\ref{sec:likelyrgb}) may bias the resulting metallicity distribution. The sample mean $\mu_\mathrm{sample}$ and sample standard deviation $\sigma_\mathrm{sample}$ of the stars with $y_0<19$\,mag are reported in the upper panels of Fig.~\ref{fig:metallicityhistograms}. We note that our observed distributions, primarily due to photometric errors,  extend beyond the [Fe/H] range where the metallicity relations are formally valid, since these are calibrated only for [Fe/H]$\ge$-2 \citep[Eq.~\ref{eq:photmetHilker};][]{2000A&A...355..994H}, [Fe/H]$\ga-3.3$ \citep[Eq.~\ref{eq:photmetAdenThesis};][]{AdenThesis}, and [Fe/H]$\ge$-2.6\,dex \citep[Eqs.~\ref{eq:photmetCalamida07vy} and \ref{eq:photmetCalamida07uy};][]{2007ApJ...670..400C}, respectively. 
 
\subsection{Intrinsic metallicity distribution function}
\label{sec:intrinsicmdf}
As mentioned in the introduction, the recent literature is mildly contradictory on the question of a spread in [Fe/H] in NGC~2419 \citep[e.g.][]{2012MNRAS.426.2889M,2013ApJ...778L..13L}. It is therefore interesting to quantify a possible spread in the photometric [Fe/H]. 

For this, we assume the intrinsic [Fe/H] distribution function of 114 clean RGB stars with $y_0<19$\,mag to be Gaussian with mean $\mu_\mathrm{intr}$ and variance $\sigma_\mathrm{intr}^2$,
\begin{equation}
p_\mathrm{intr}\left(\FeH\given\muintr,\sigintr\right) = \frac{1}{\sigintr \sqrt{2\pi}}\mathrm{exp}\left[\frac{-\left(\mathrm{[Fe/H]}-\muintr\right)^2}{2\sigintr^2}\right].
\label{eq:intrinsicmdf}
\end{equation}

Our measured values differ from the actual values by a measurement error $x=\FeHobs-\FeH$, which we assume to be drawn from an uncertainty distribution $\epsilon\left(x, \param_i\right)$ that remains to be specified and that is parametrised by a vector $\param_i$ depending on the star $i$, because the photometric uncertainty varies stellar magnitude, and because  crowding also varies with the location on the sky.

The likelihood is then 
\begin{equation}
\label{eq:likelihood}
\begin{split}
\likelihood&\left(\FeHobsi\given\muintr,\sigintr,\param_i\right) = \\
&\prod_{i} \left(p_\mathrm{intr}\ast\epsilon_i\right)\left(\FeHobsi\given\muintr,\sigintr,\param_i\right),
\end{split}
\end{equation}
where $\ast$ denotes convolution. Ultimately, we are interested in the posterior probability distribution function (pdf) of the parameters $\muintr$ and $\sigintr$ of the cluster's intrinsic metallicity distribution that according to Bayes' theorem is
\begin{equation}
\begin{split}
p&\left(\muintr,\sigintr\given\FeHobsi,\theta_i\right) =\\ &\frac{\likelihood\left(\FeHobsi\given\muintr,\sigintr,\param_i\right)~p\left(\muintr,\sigintr\right)}{p\left(\FeHobsi\right)}.
\end{split}
\label{eq:posteriorpdf}
\end{equation} 
Here $p\left(\muintr,\sigintr\right)$ is the prior pdf for $\muintr$ and $\sigintr$ and the denominator $p\left(\FeHobsi\right)$ is the marginal likelihood that for a given model is a constant that we avoid calculating as it does not affect our parameter inference.

\begin{figure}
\includegraphics[width=\linewidth]{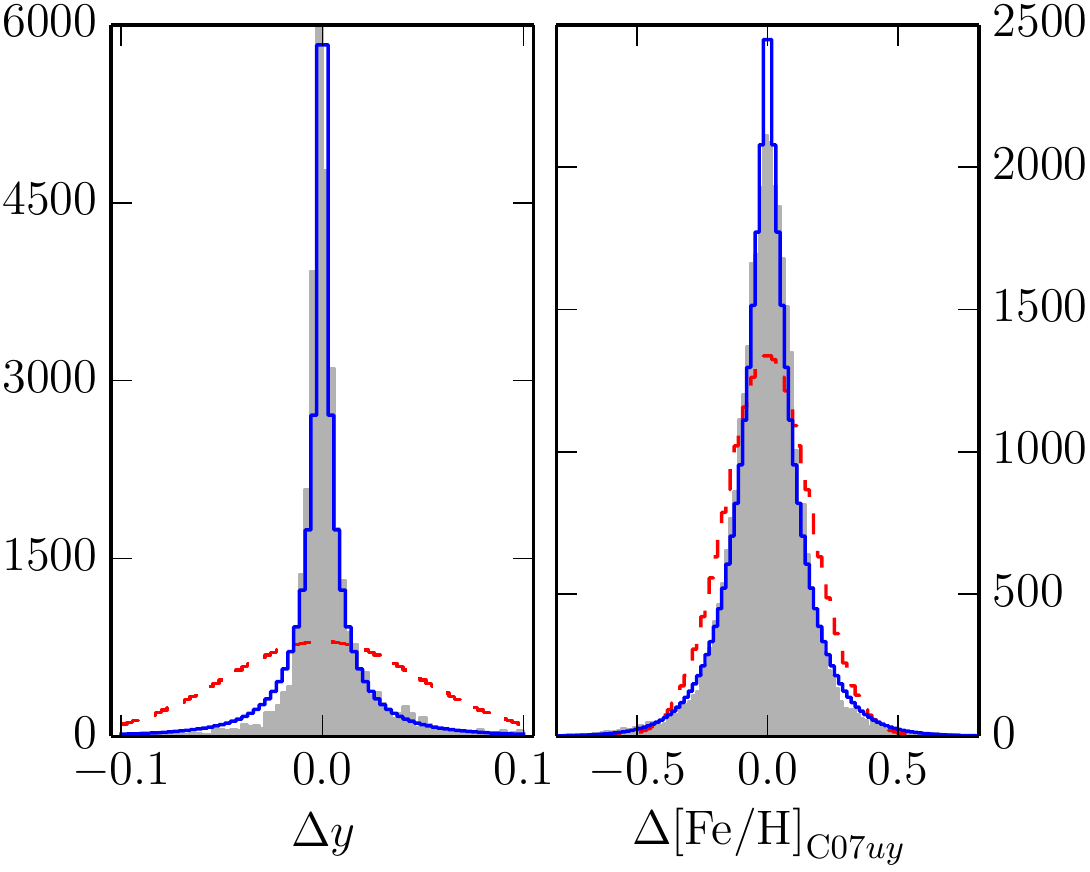}
\caption{Grey histograms show the distribution of uncertainties in $y$ (left panel) and photometric metallicity $\mathrm{[Fe/H]}_{\mathrm{C07}uy}$ (Eq.~\ref{eq:photmetCalamida07uy}, right panel), as derived from artificial star tests for stars with $18<y<19$\,mag. The red dashed curves shows the maximum-likelihood zero-mean normal distribution, the blue solid curves correspond to the maximum-likelihood zero-mean generalised Gaussian distribution (Eq.~\ref{eq:gengaussian}).}
\label{fig:errormodel}
\end{figure}

In the case of measurement errors distributed normally around zero mean with variances $\sigma_{\mathrm{obs},i}^2$, the convolution in Eq.~\ref{eq:likelihood} yields the usual analytic expression, i.e. a product of $i$ Gaussians with variances $\left(\sigintr^2+\sigma_{\mathrm{obs},i}^2\right)$ \citep[e.g.][]{1993ASPC...50..357P}. However, for our data, the assumption of Gaussian uncertainties turned out to be a poor one. To see this, we estimate the distribution of measurement errors $\epsilon_\mathrm{i}$ empirically from the artificial star test results in the following way. For each observed star, we select successfully recovered artificial stars that have an inserted magnitude within 0.25\,mag of the star's observed magnitude in all four filters and that are at a similar radial distance (within 10\,arcsec) from the cluster centre. This typically yields a few tens to a few hundreds of artificial stars (with a median number of 220). The uncertainty distributions in a given magnitude or colour can then be estimated directly from the deviations of inserted and measured magnitudes, e.g. $\Delta y=y_\mathrm{in} - y_\mathrm{out}$. The uncertainties in the photometric metallicities also depend on the \emph{nominal} values of $m_1$ and the colour index and therefore can be estimated by \emph{varying} the observed star's magnitudes by the magnitude deviations found for a given artificial star, e.g. $y_\mathrm{obs} + \Delta y$, calculating the photometric metallicities using these varied magnitudes, and subtracting the observed star's nominal photometric metallicity.

Examining histograms of these empirical uncertainties shows that their distribution deviates significantly from the Gaussian shape. This is illustrated in Fig.~\ref{fig:errormodel}, which shows  the distributions of deviations in $y$ and the photometric metallicity according to Eq.~\ref{eq:photmetCalamida07uy}; for clarity we increased the sample size by showing the deviations obtained for all stars $18<y_0<19$\,mag, rather than for a single star. Compared to the maximum-likelihood Gaussians (shown as red dashed curves) the histograms have more acute peaks and more extended tails.
While it is clear that the common assumption of Gaussian uncertainties virtually never holds in detail in any set of observational data, it is intuitively plausible that the rather strong deviations from normality in our data are due in  large part to the mediocre seeing during the observations. In the worst-seeing images, the PSF FWHM is $\sim\!$3\arcsec or 9 pixels, increasing the chance of, for example,   a detector defect, hot pixel, cosmic ray, or contamination from an imperfectly subtracted neighbour  falling within the PSF-fitting radius and producing `outlier' measurements that are responsible for the long tails. By assuming normally distributed uncertainties, we would therefore overestimate the typical photometric uncertainty of a star, leading to an underestimation of the intrinsic metallicity dispersion. We therefore discard this assumption in favour of an error model of the `generalised Gaussian' form \citep[e.g.][]{VaranasiThesis} that allows for a variation of the distribution's kurtosis,
\begin{equation}
f_\mathrm{genG}\left(x, \mu, \sigma, \beta \right) = \frac{1}{2\Gamma\left(1+1/\beta\right)\alpha\left(\sigma,\beta\right)} \mathrm{exp}\left[-\left(\frac{|{x-\mu}|}{\alpha\left(\sigma,\beta\right)}\right)^{\beta}\right],
\label{eq:gengaussian}
\end{equation} with $\alpha\left(\sigma,\beta\right) =\sqrt{\sigma^2\Gamma\left(1/\beta\right)/\Gamma\left(3/\beta\right)}$, where $\sigma^2$ is the variance, $\Gamma$ is the $\Gamma$-function, and the exponent $\beta$ determines the shape. For $\beta=2$, Eq.~\ref{eq:gengaussian} becomes the Gaussian distribution. This error model, while not being a perfect fit, much more accurately reproduces the empirical uncertainty distributions, as can be seen in Fig.~\ref{fig:errormodel} where the solid blue curves represent the maximum-likelihood generalised Gaussians. The more realistic error model, however, has the disadvantage of increasing computational cost, since the convolution integrals in Eq.\ref{eq:likelihood} no longer have a simple analytic form.

With this, we set the uncertainty distribution for each star~$i$ in Eq.~\ref{eq:likelihood} to the zero-mean generalised Gaussian $\epsilon_i\left(x, \param_i\right) = f_\mathrm{genG}\left(x, 0, \hat\sigma_i, \hat\beta_i\right)$, where $\hat\sigma_i$ and $\hat\beta_i$ are the maximum-likelihood point-estimate obtained from the empirical uncertainty distribution of artificial stars that are close to  that star in magnitudes and radius in the sense described above. With the error model specified, we can choose a prior \citep[we use the non-informative prior that is uniform in $\muintr$ and in $\mathrm{log}\,\sigintr$, following the invariance argument of][]{JeffreysPrior} and sample from the posterior pdf (Eq.~\ref{eq:posteriorpdf}). In practice, we used the \textsc{emcee} Markov chain Monte Carlo (MCMC) sampler \citep{2013PASP..125..306F} in combination with a \textsc{Cython} \citep{behnel2010cython} interface to the \textsc{qagi} numerical integrator of the Gnu Scientific Library \citep{galassi2009gnu} in order to evaluate the convolutions in Eq.~\ref{eq:likelihood}.   

The resulting intrinsic dispersions $\sigintr$ and means $\muintr$ are displayed in the upper panels of Fig.~\ref{fig:metallicityhistograms}, where the nominal values correspond to the median of the posterior pdf of each parameter and the uncertainties correspond to the 1-$\sigma$ equivalent (68\,per cent) credible intervals. For all but the \citet{AdenThesis} metallicity relation, we obtain small intrinsic spreads of the photometric metallicity of $\sigintr\sim$0.11-0.15\,dex. These spreads are formally highly significant (i.e. different from zero), but they are also on the level of the internal precision expected for the photometric metallicity estimation; \citet{2000A&A...355..994H} and \citet{2007ApJ...670..400C} report a residual scatter in the photometric metallicity of their \emph{calibrating} stars of $\sigma=0.16$ and $0.21$\,dex, respectively, although in both cases, these are due to the combined effects of systematics (e.g. due to different photometric calibrations and reddening uncertainties), photometric errors, and  the intrinsic accuracy of the photometric metallicity estimation. In our data, we also see systematic tilts of the sequence of observed RGB stars compared to the iso-metallicity curves in Fig.~\ref{fig:metallicityrelations}, which lead to an unphysical trend of photometric [Fe/H] with magnitude and artificially increase the spread in [Fe/H]. This tilt is less pronounced, although still visible, in the \citet{2000A&A...355..994H} metallicity relation (Eq.~\ref{eq:photmetHilker}), so that we adopt $\sigintr=0.11^{+0.02}_{-0.01}$\,dex as the best estimate of the intrinsic spread. 

Furthermore, as we will see in Section~\ref{sec:multiplepop}, the apparent spread in nitrogen abundance in NGC~2419 causes a spread in $m_{1,0}$ of the order of $\sim$0.025\,mag, using e.g. Eq.\ref{eq:photmetHilker}, which is valid only for CN-weak first-generation stars. This translates to an artificial spread in $\sim$0.1\,dex in photometric [Fe/H] for a bright red giant. 

Thus our derived formal [Fe/H] spread of $\sigintr=0.11^{+0.02}_{-0.01}$\,dex is an upper limit of the true metallicity spread, and is at least partially (but possibly entirely) caused by the limited accuracy of the photometric metallicity measurement and the presence of systematic effects, suggesting that there may be very little room for a substantial true intrinsic [Fe/H] spread in NGC~2419.

\subsection{Comparison with spectroscopic metallicity}
Recent spectroscopic studies found average iron abundances for NGC~2419 of 
[Fe/H]=$-2.06$\,dex \citep{2011ApJ...740...60C} and [Fe/H]=$-2.09\pm0.02$\,dex (\citealt{2012MNRAS.426.2889M} and \citealt{2012ApJ...760...86C}). In the previous section we derived mean intrinsic [Fe/H] of $\mu_\mathrm{intr,A11}=-2.43^{+0.03}_{-0.02}$, $\mu_\mathrm{intr,H00}=-1.96\pm0.01$, $\mu_\mathrm{intr,C07vy}=-2.07\pm0.02$, and $\mu_\mathrm{intr,C07uy}=-2.08\pm0.02$\,dex. All but the first of these are within 0.13\,dex of the spectroscopic value of $-2.09\pm0.02$\,dex, with the last two estimates agreeing with the spectroscopic value within the errors. This agreement is reassuring and suggests that, despite the differences between the individual relations that are apparent in Fig.~\ref{fig:metallicityrelations}, most existing \Sgren metallicity calibrations provide accurate and consistent metallicity estimates for a population of metal-poor red giants.

To compare photometric and spectroscopic [Fe/H] for individual stars, 
we cross-matched our clean RGB stars catalogue with the spectroscopic samples of \citet{2012MNRAS.426.2889M} and \citet{2012ApJ...760...86C}, which yielded 27 and  8 matches with spectroscopic [Fe/H] measurements, respectively. For the four stars that the two samples have  in common, we averaged the spectroscopic [Fe/H] from the two studies. For the resulting 31 matches, the lower panels of Fig.~\ref{fig:metallicityhistograms} show the spectroscopic [Fe/H] plotted against the photometric [Fe/H] according to the four metallicity relations. We note that the errorbars on the photometric values in this diagram represent only the statistical uncertainties, i.e. the magnitude uncertainties propagated through Eqs.~\ref{eq:photmetHilker}~to~\ref{eq:photmetAdenThesis}, taking into account the uncertainties in the calibration coefficients where available. These are naturally smaller than the uncertainties derived from artificial star tests in the previous section, which also take into account the imperfections of photometry (e.g. imperfectly subtracted neighbours, blends, bad pixels).

\begin{figure}
\includegraphics[width=\linewidth]{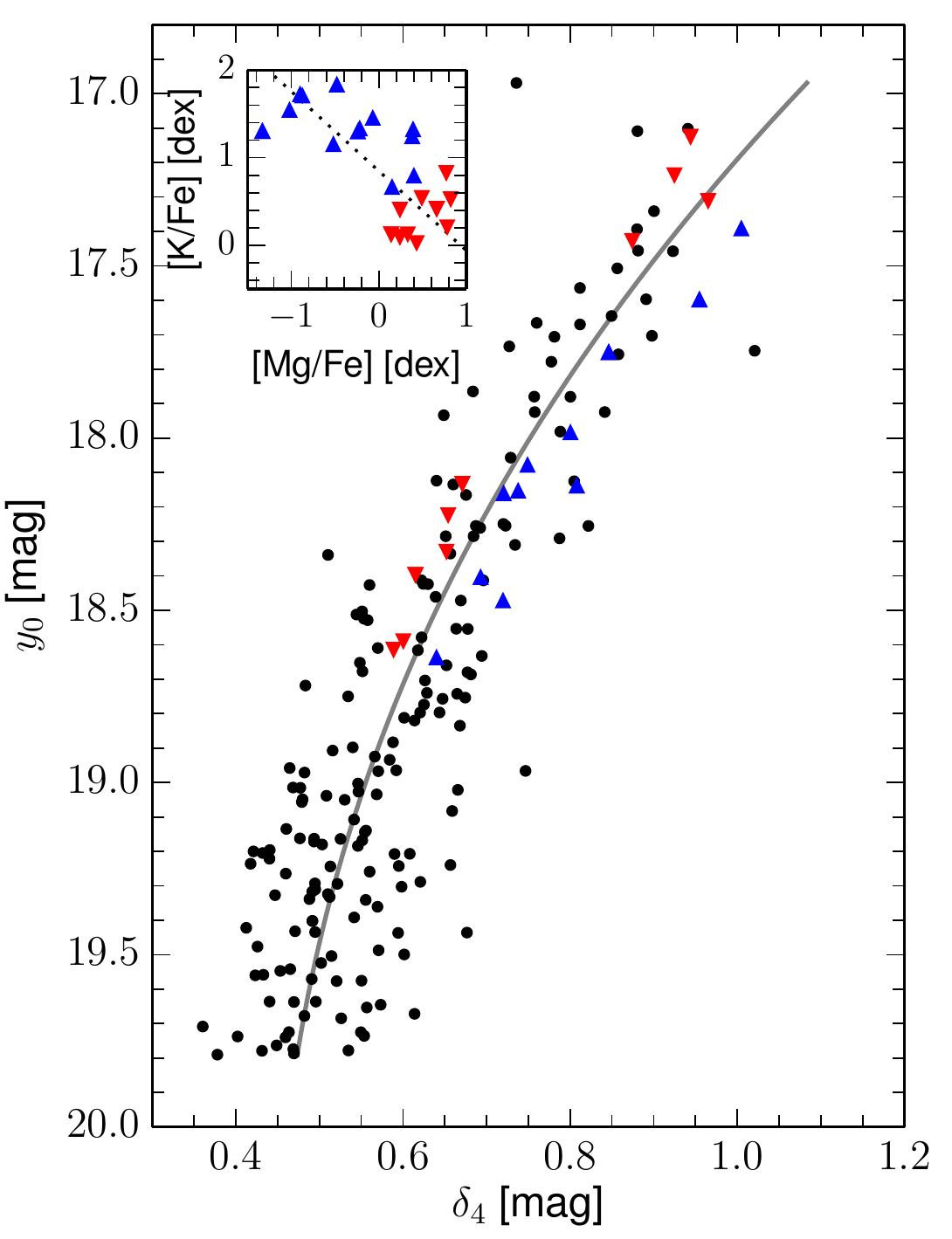}
\caption{$y_0$ vs. $\delta_4$ colour-magnitude diagram of stars in the clean RGB sample. Triangles correspond to stars with [Mg/Fe] and [K/Fe] measured by \citet{2012ApJ...760...86C} and/or \citet{2012MNRAS.426.2889M}. These abundances are shown in the inset, and we divided the sample into a Mg-poor, K-rich (blue $\blacktriangle$), and Mg-rich, K-poor (red $\blacktriangledown$) group. The ridge line, shown as a grey curve, is a third-order polynomial in $\delta_4$ as a function of $y_0$. All stars in the Mg-poor, K-rich group lie on the red side of the RGB in $\delta_4$, just like the Na- (and N-) rich second-generation stars in other globular clusters.} 
\label{fig:spreadCMDd40}
\end{figure}

On a star-by-star basis, the semi-empirical \citet{2007ApJ...670..400C} metallicity relations are in excellent agreement with spectroscopy. The \citet{2000A&A...355..994H} relation yields consistent measurements that are systematically biased towards high metallicities by $\sim\!0.2$\,dex, while the \citet{AdenThesis} relation produces similarly large offset towards low metallicities and a larger scatter. Overall, the semi-empirical \citet{2007ApJ...670..400C} metallicity relations best reproduce the individual and the mean [Fe/H] of RGB stars in NGC~2419 and may be the best choice among the three studies for deriving the [Fe/H] of giants from \Sgren photometry.

The comparison between spectroscopic and photometric metallicities also provides a further clue against the existence of a significant spread in [Fe/H]. The sample standard deviations of the spectroscopic ($\sim\!0.1$\,dex) and photometric metallicities ($\sim\!0.15$\,dex, except for those obtained using the \citet{AdenThesis} calibration) are comparable, indicating a similar precision of the measurement. Thus, if there were a physical spread in [Fe/H] of similar magnitude, one expects the spectroscopic and photometric metallicity to be correlated. However, the linear or Pearson's  correlation coefficients $\rho$ (see lower panels of Fig.~\ref{fig:metallicityhistograms}) suggest that the two quantities are only weakly correlated, with $\rho=0.2$--$0.3$. The probability for achieving a correlation at least as high by chance ($p_\mathrm{rand}$) is 13--26~per cent, i.e. the weak correlations are less significant than at the $2\sigma$ level. This lack of  significant correlation supports the interpretation that there is no substantial internal [Fe/H] spread in NGC~2419.

\begin{figure*}
\includegraphics[width=0.33\linewidth]{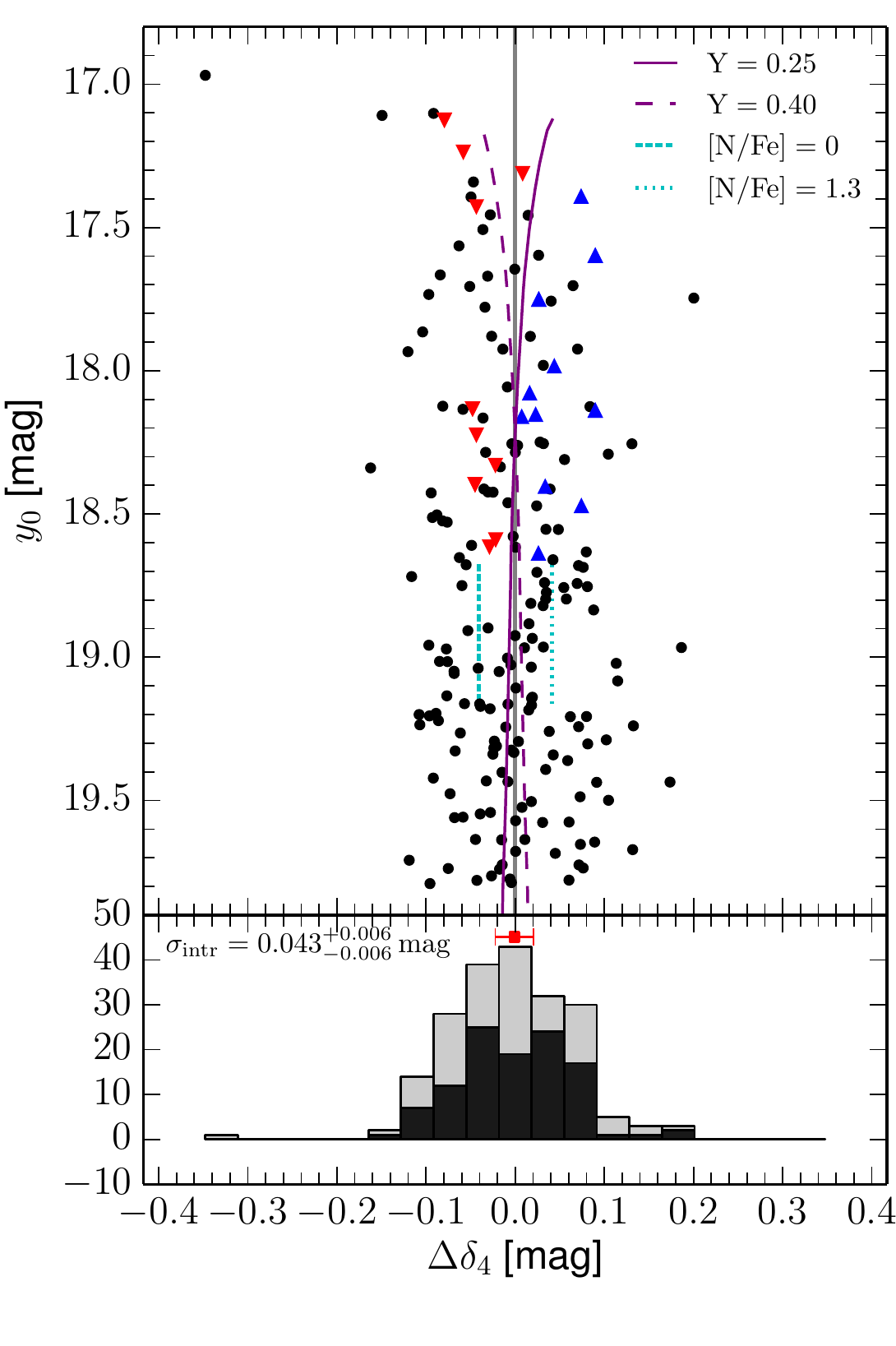}
\includegraphics[width=0.33\linewidth]{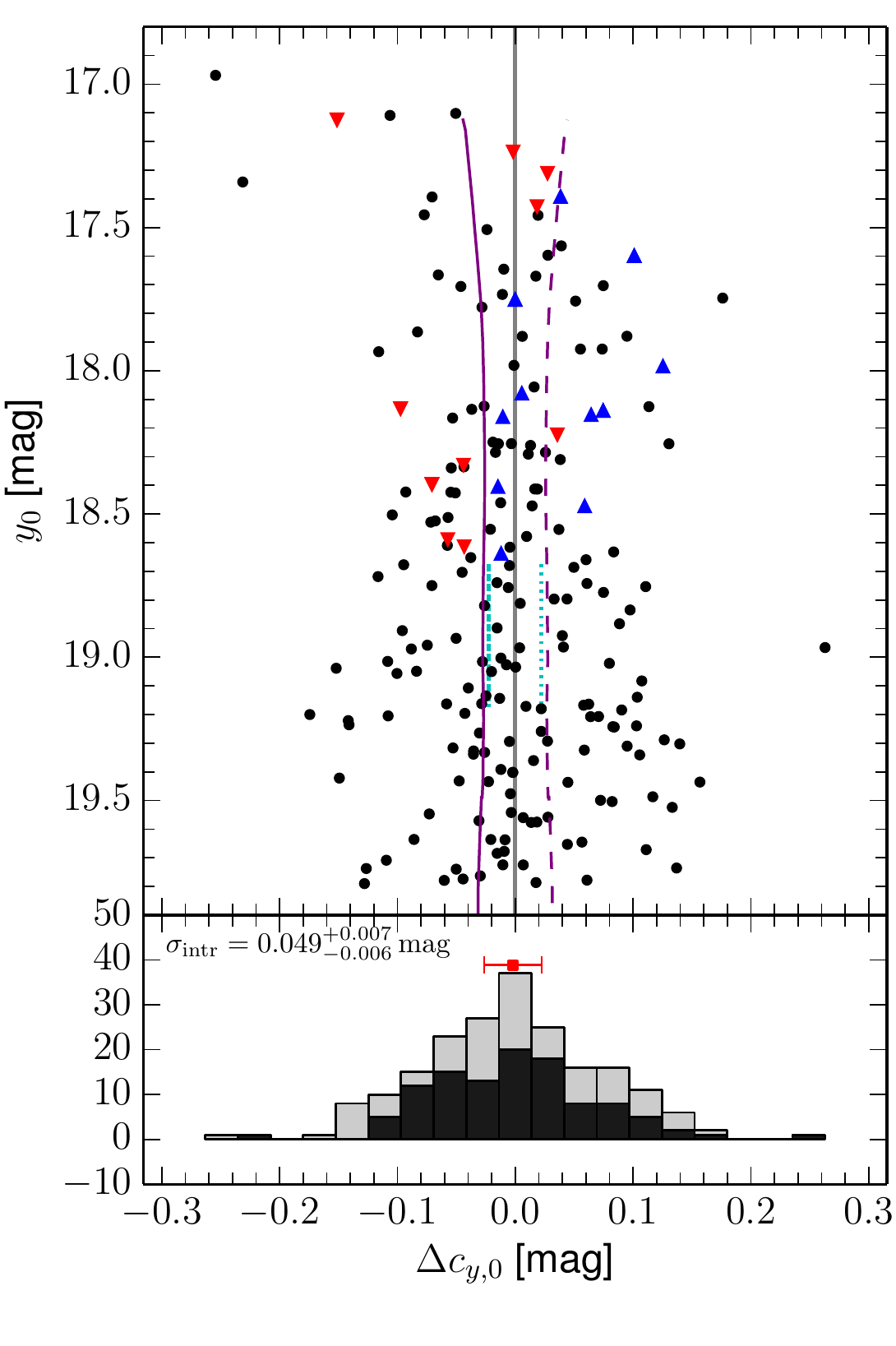}
\includegraphics[width=0.33\linewidth]{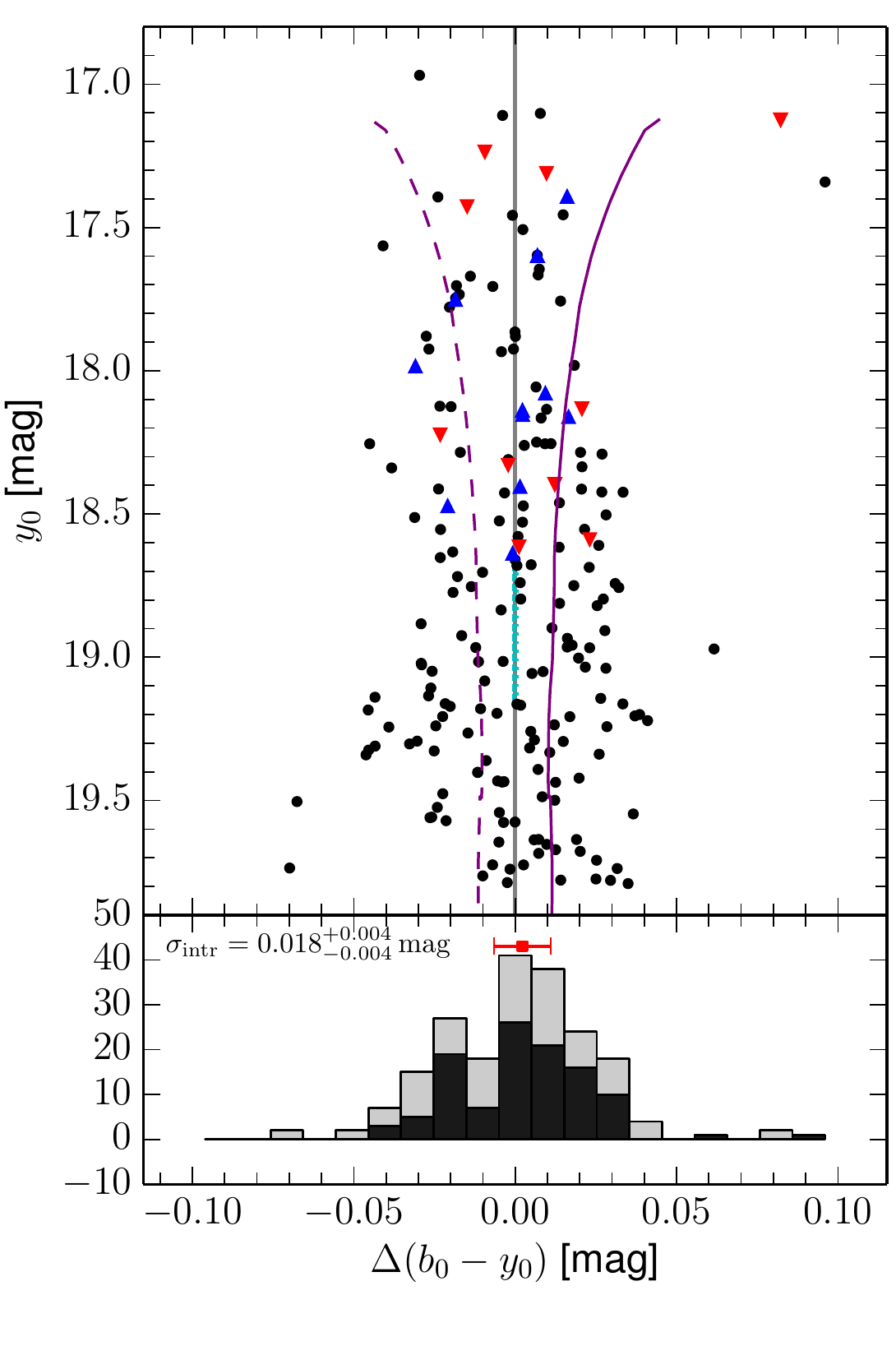}
\includegraphics[width=0.33\linewidth]{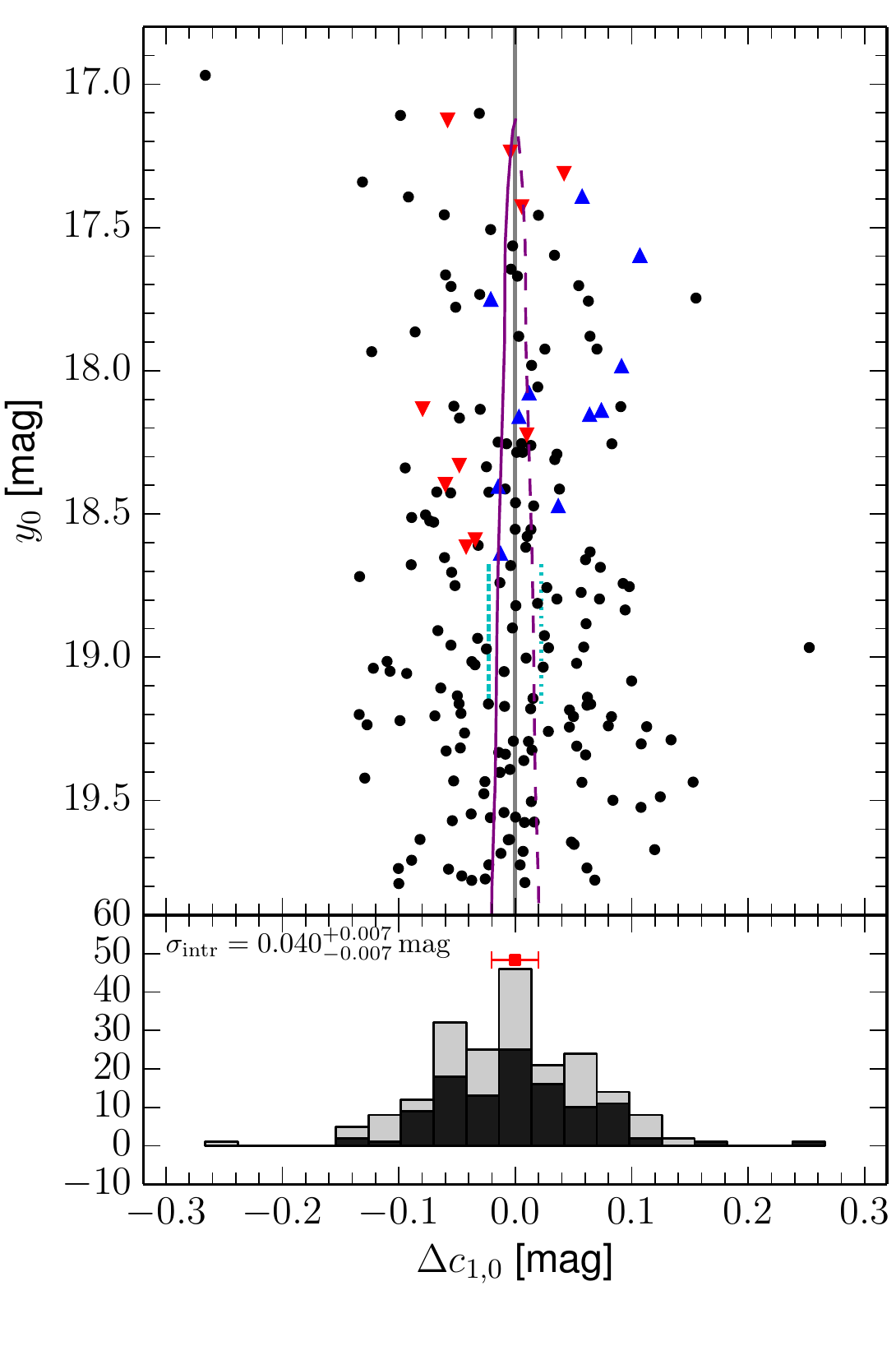}
\includegraphics[width=0.33\linewidth]{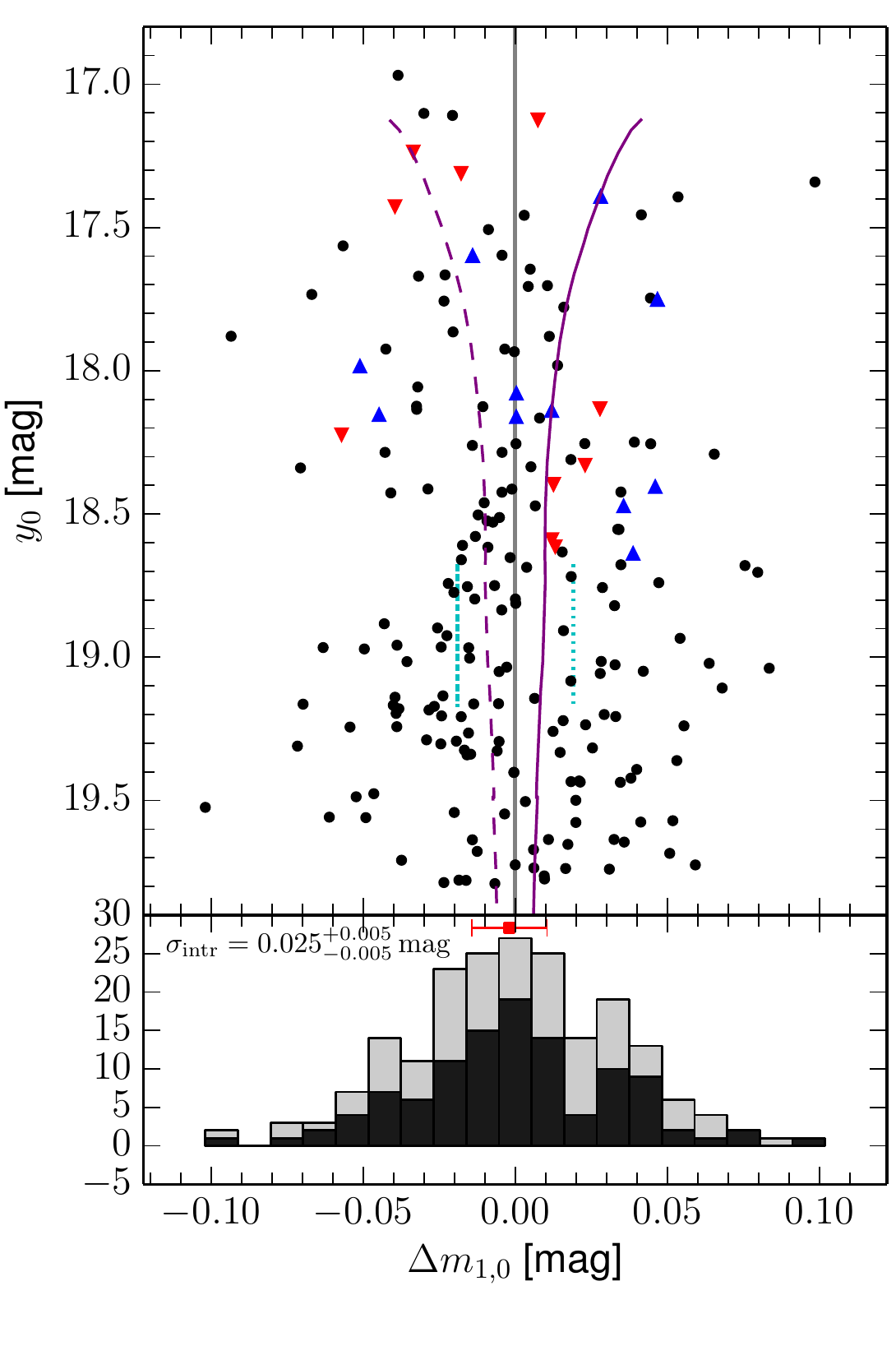}
\includegraphics[width=0.33\linewidth]{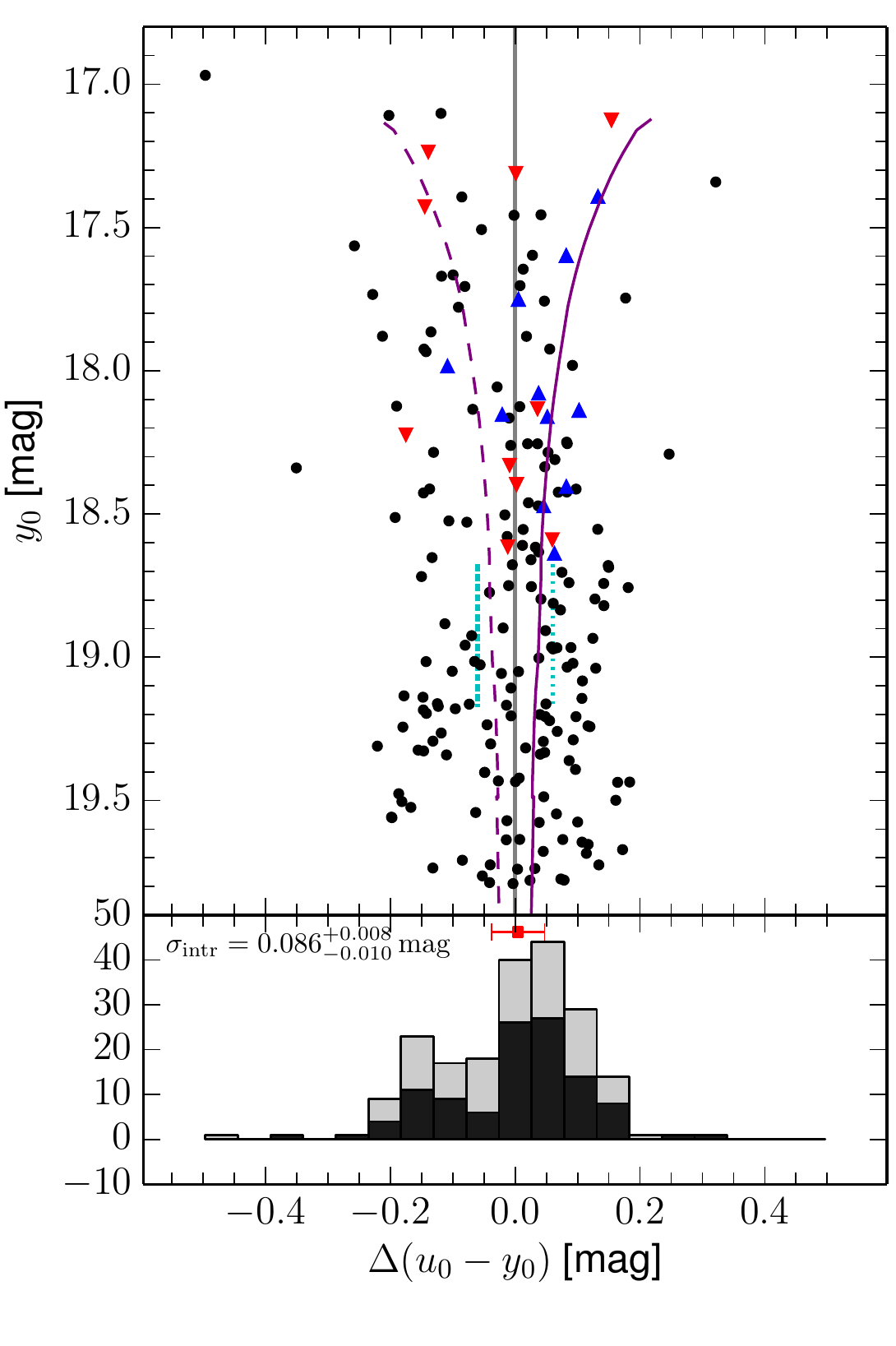}
\caption{{\bf Upper panels} show colour magnitude diagrams of $y_0$ vs. $\delta_4$, $c_{y,0}$, $(b-y)_{0}$, $c_{1,0}$, $m_0$ and $(u-y)_0$, respectively, where the colours were detrended by subtracting a ridge line polynomial (for $\delta_4$ this is shown in Fig.~\ref{fig:spreadCMDd40}). Symbols as in Fig.~\ref{fig:spreadCMDd40}. The purple lines demonstrate the differential effect of He enhancement by $\Delta$Y=0.15. They correspond to 13 Gyr, [Fe/H]=-2\,dex, [$\alpha$/Fe]=+0.4\,dex Dartmouth isochrones, with Y=0.25 (solid purple line) and Y=0.40 (dashed purple line), respectively, from which the average colour at a given magnitude has been subtracted. The cyan vertical lines around $y_0$=18.3\,mag show the expected difference between C-normal, N-poor (dashed line) and C-normal, N-rich stars (dotted line) with M$_V=-1$\,mag taken from \citet{2011A&A...535A.121C}. {\bf Lower panels} show histograms of the colour distribution of all stars in the clean RGB sample (light histograms) and of stars with $17.2<y_0<19$\,mag (dark histograms). The intrinsic spreads $\sigma_\mathrm{intr}$ obtained by assuming a Gaussian intrinsic distribution are reported and illustrated by the red errorbars above the histograms.} 
\label{fig:spreadd40cybmy}
\end{figure*}

\section{Nitrogen and He variations}
\label{sec:multiplepop}
The \Sgren $v$ and particularly $u$ bands are known to be sensitive to variations in carbon and nitrogen. \citet{2008ApJ...684.1159Y} found that the nitrogen abundance of red giants in NGC~6752 correlates with $c_1$. They defined a new index, $c_y= c_1- (b-y)$, which to first order removed the dependency of $c_1$ on effective temperature, and found a direct proportionality between $c_y$ and [N/Fe]. \citet{2011A&A...535A.121C} similarly defined a new index, $\delta_4=(u-v)-(b-y)$, which in a sample of GCs maximised the separation between first- and second-generation stars that were spectroscopically identified via their Na and O abundances. 

For NGC~2419, the existence of at least two chemically different subpopulations, the Mg-rich/K-poor and the extreme Mg-depleted/K-rich populations, has been found spectroscopically \citep{2012MNRAS.426.2889M,2012ApJ...760...86C}. It also has been shown that the two populations separate in the \Sgren $hk=(Ca-b)-(b-y)$ index, where $Ca$ is a narrowband filter encompassing the Ca H and K lines \citep{2013ApJ...778L..13L}, which supports a variation in Ca between the two populations. The two populations also separate in broadband $u_\mathrm{SDSS}-V$ colour \citep{2013MNRAS.431.1995B}. These authors attribute the observed colour spreads to a difference in the He abundance, but note that the Mg-deficient presumably second-generation stars, puzzlingly, have redder $u_\mathrm{SDSS}-V$ colours, while a second generation enhanced in He would have bluer colours than the first generation. 

Here, we show that the two populations also separate in the CN-sensitive \Sgren indices. Figure~\ref{fig:spreadCMDd40} shows the $y_0$ magnitude vs. $\delta_{4,0}$ index of stars in the clean RGB sample. Stars with spectroscopic Mg and K abundances from \citet{2012ApJ...760...86C} and \citet{2012MNRAS.426.2889M} are highlighted as coloured triangles; as before, we averaged the element abundances from the two studies for the four stars in common. We divided the spectroscopic sample into a Mg-rich/K-poor (red $\blacktriangledown$ symbols) and a Mg-poor/K-rich (blue $\blacktriangle$ symbols) group by splitting the sample at roughly the median value of the quantity $0.74($[Mg/Fe]-$<$[Mg/Fe]$>)-0.68($[K/Fe]$-<$[K/Fe]$>)$. This quantity reflects the first principal component of the [Mg/Fe], [K/Fe] distribution, with (0.74,-0.68) being the vector along the direction of the anti-correlation that is shown as a dotted line in the inset in Fig.~\ref{fig:spreadCMDd40}. The two populations separate clearly in this CMD, with the Mg-poor/K-rich (presumably second-generation stars) lying on the red side of the RGB, just as the Na-rich second-generation stars in other GCs have redder $\delta_4$ colours \citep{2011A&A...535A.121C}.

In order to remove the temperature-dependence of $\delta_4$, we subtracted the colour of a ridge line, shown as the grey curve in Fig.~\ref{fig:spreadCMDd40}, from the data to obtain a detrended $\Delta\delta_4$. The ridge line is simply taken as a third-order polynomial, taking colour to be a function of $y_0$ magnitude (i.e. ignoring the smaller uncertainties on $y_0$) and using an iteratively reweighted least squares minimisation \citep{BeatonTukey1974} in order to be less sensitive to outliers. The $\Delta\delta_{4,0}-y_0$ diagram is shown in the upper left panel of Fig.~\ref{fig:spreadd40cybmy}. The remaining panels show the corresponding ridge-line-detrended CMDs with $\Delta c_{y,0}$ and $\Delta(b-y)_0$, $\Delta c_{1,0}$, $\Delta m_{1,0}$, and $\Delta(u-y)_0$ on the colour-axis. It can be seen that the two spectroscopically identified subpopulations separate most clearly in $\delta_4$, while  $c_y$ and $c_1$ also show some degree of segregation. In $(b-y)$ and $m_1$ no clear separation is seen. Surprisingly, there is also no segregation in $(u-y)_0$ even though this colour is expected to depend strongly on variations in CN. To see this, cyan lines in the upper panels represent the expected difference, interpolated to a metallicity of [Fe/H]=-2\,dex, between N-rich (dotted line) and N-poor (dashed line) stars at M$_V$=-1\,mag ($y_0\sim\!$18.9\,mag) according to the spectral synthesis models by \citet{2011A&A...535A.121C}. These authors calculated the impact of light-element variations on \Sgren photometry for a single enrichment pattern assuming [N/Fe]=[C/Fe]=0\,dex for the first-generation, N-poor stars and [N/Fe]=+1.3\,dex and [C/Fe]=-0.2 dex for the second-generation, N-rich stars (see their paper for details and the complete assumed abundance patterns).

The histograms in the lower subpanels of Fig.~\ref{fig:spreadd40cybmy} show the distribution of all clean RGB stars in the detrended colour (light histograms) and that  only of stars brighter than $y_0=19$\,mag and list the intrinsic dispersions in the brighter sample, which were obtained using the same formalism as in Section~\ref{sec:photmet}, i.e. by assuming the intrinsic distribution to be Gaussian and modelling the uncertainties in each colour or index with a generalised Gaussian distribution fit to the artificial star test results. 
The magnitude of the spreads, 0.04-0.05\,mag in $\delta_4$ and $c_y$, is comparable to that seen in other metal-poor clusters: M92 ([Fe/H]$<$2~dex, \citealt{1999PASP..111.1115S}) shows a spread of $\delta_4\sim0.05$\,mag \citep{2011A&A...535A.121C}. NGC 6397, M15, and M92 show spreads in $c_y$ of 0.05 to 0.1\,mag \citep{2008ApJ...684.1159Y}.

\begin{figure}
\includegraphics[width=\linewidth]{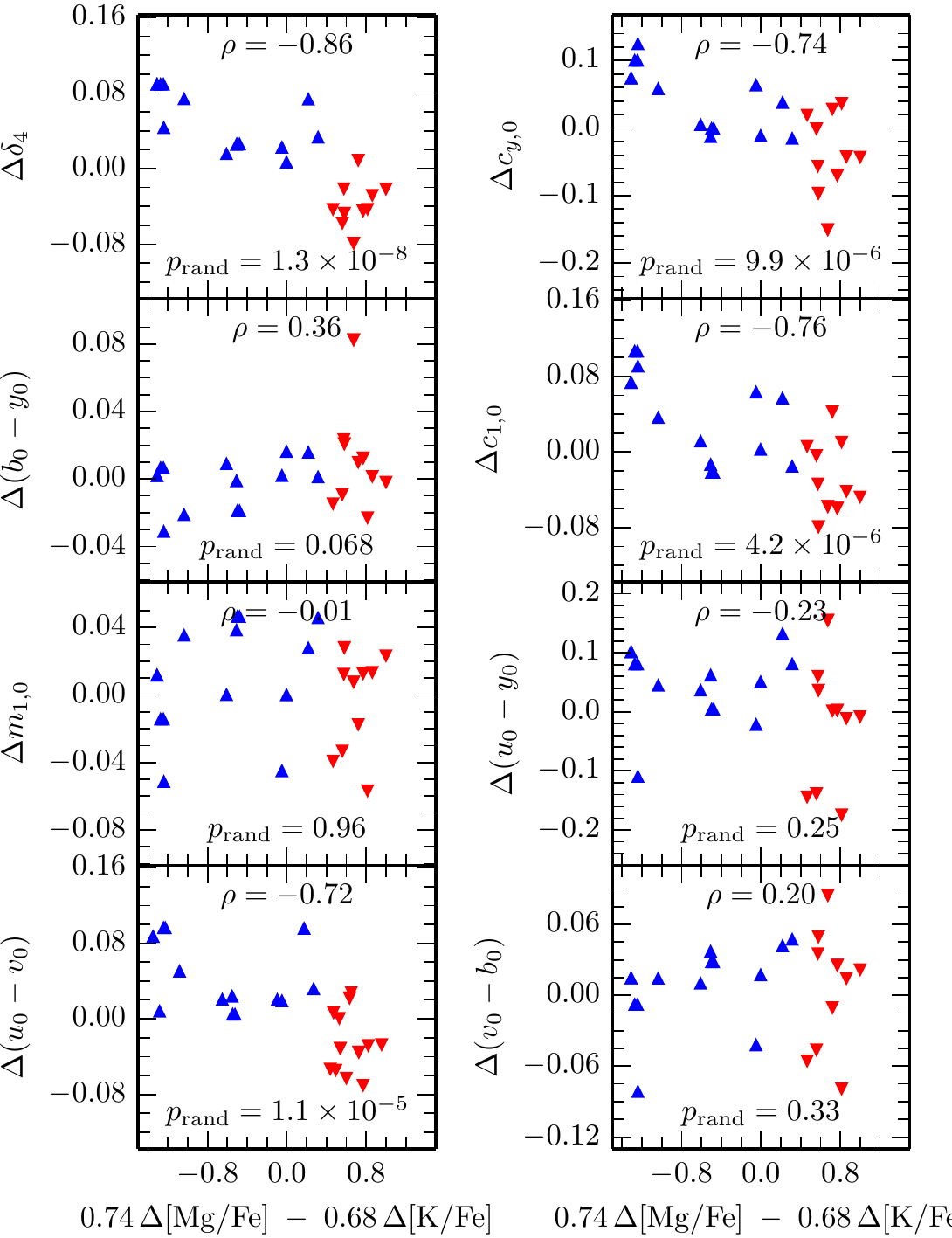}
\caption{Correlations of ridge-line-detrended colours on the RGB with spectroscopic with $0.74($[Mg/Fe]-$<$[Mg/Fe]$>)-0.68($[K/Fe]$-<$[K/Fe]$>)$, the first principal component of the [Mg/Fe], [K/Fe] distribution. Linear correlation coefficients $\rho$ and probabilities for the correlation to be by chance $p_\mathrm{rand}$ are reported in each panel.}
\label{fig:correlations}
\end{figure}

The correlations of the various ridge-line-detrended colours with principal component of the Mg-K distribution are shown in Fig.\ref{fig:correlations}. Where there is a significant correlation ($\delta_4$, $c_1$, $c_y$, and $(u-v)_0$), it is generally strongest and most significant with the principal component projection, somewhat weaker with [Mg/Fe], and much weaker and in the opposite direction with [K/Fe]. The $(b-y)_0$ colour correlates with the abundances, although at lower significance, and in the opposite sense than the other colours and indices. Again, there is little correlation between Mg/K abundance and $(u-y)_0$.

The key to understanding these patterns is to also take the variation of He into account, which (depending on the colour index) can act in the opposite sense of nitrogen enrichment \citep{2011A&A...534A...9S}. The expected differential effect of an enhancement in He of $\Delta$Y=0.15, which is somewhat more extreme than the $\Delta$Y$\sim$\!0.1 favoured for NGC~2419 \citep{2015MNRAS.446.1469D}, is shown by the purple curves in the CMDs in Fig.~\ref{fig:spreadd40cybmy}. These curves were obtained by subtracting the average colour at a given magnitude from each of two isochrones (one with Y=0.25, one with Y=0.40) with 13~Gyr, [Fe/H]=-2\,dex, and [$\alpha$/Fe]=+0.4\,dex. Comparison of the observational scatter with the spread of these two models gives an indication of whether He plays a role for a given colour index. The parameters $\delta_4$ and $c_1$ are essentially independent of He; $c_y$ shows some dependence on He content, but here the effect is aligned with that of a nitrogen variation: enhancements in both elements lead to redder colours. The weak correlation of $(b-y)_0$ with Mg and K abundances in the opposite direction compared to all other colour indices can also be explained by He enhancement, since $(b-y)_0$ is independent of nitrogen \citep{2011A&A...535A.121C}, but shows a moderate dependence on He content. Finally,  the lack of correlation in $(u-y)_0$ and $m_{1,0}$ in Fig.~\ref{fig:correlations}, as well as the very strong correlation in $(u-v)_0$ can also be attributed to He enrichment: $(u-y)_0$ and $m_{1,0}$ are sensitive to He content, and the effect of He enrichment acts against that of nitrogen enrichment; $(u-v)_0$ on the other hand is much less sensitive to He content, but shows similar sensitivity to nitrogen.

The internal spread in $\delta_4$, which is least influenced by He abundance, is roughly half the value  predicted by \citet{2011A&A...535A.121C}. This may indicate that the N-rich second population has [N/Fe] lower than 1.3\,dex, although for a quantitative analysis, models for more than a single enrichment value would be required. Also, because the deficiency of the electron-donor element Mg in NGC~2419's second generation can have significant impact on the continuous opacity as shown by \citet{2012MNRAS.426.2889M}, spectral synthesis models calculated specifically for the observed abundance pattern in NGC~2419, preferably also taking  the effect of element variations on stellar structure into account \citep[e.g.][]{2011A&A...534A...9S} would be highly desirable for a quantitative statement on the [N/Fe] abundance.

\subsection{A bimodality in $\delta_4$?}
Given the ongoing efforts to characterise multiple populations in GCs using \Sgren photometry \citep[e.g.][]{2013MmSAI..84...63A}, it is interesting to test, whether one can infer a bimodality from our photometric data, i.e. the existence of two distinct populations of stars with different nitrogen enrichment, rather than a continuous distribution. We assessed this using the distribution of $\Delta\delta_4$, because it shows the cleanest dependence on nitrogen enrichment. For this, in addition to the model assuming an intrinsic Gaussian distribution, we fit the $\Delta\delta_4$ distribution (for RGB stars with $y_0$<19\,mag) with a mixture model consisting of two $\delta$-peaks, separated by a distance $d>0$:
\begin{equation}
p_\mathrm{intr,mixture}= q\,\delta\!\left(\mu+d/2\right) + (1-q)\,\delta\!\left(\mu-d/2\right).
\label{eq:mixturemodel}
\end{equation}
Here $\delta$ is the $\delta$ function and $q$ is the relative contribution to the mixture of the population `redder' in $\delta_4$, i.e. the Mg-deficient, N-rich presumable second generation. As before, to obtain the likelihood function, Eq.~\ref{eq:mixturemodel} is convolved with the distribution of photometric uncertainties $\epsilon_i$ of the form given in Eq.~\ref{eq:gengaussian} with parameters derived from artificial stars. We chose an uninformative prior for $\mu$ and $d$ (uniform in $\mu$ and in $\mathrm{log\,}d$), and restrict $q$ to the interval $\left[0.1,0.9\right]$ in order to avoid the fit converging on a situation where one component of the mixture represents only the single star with the smallest uncertainty, thereby strongly increasing the model's overall likelihood. 

\begin{figure}
\includegraphics[width=\linewidth]{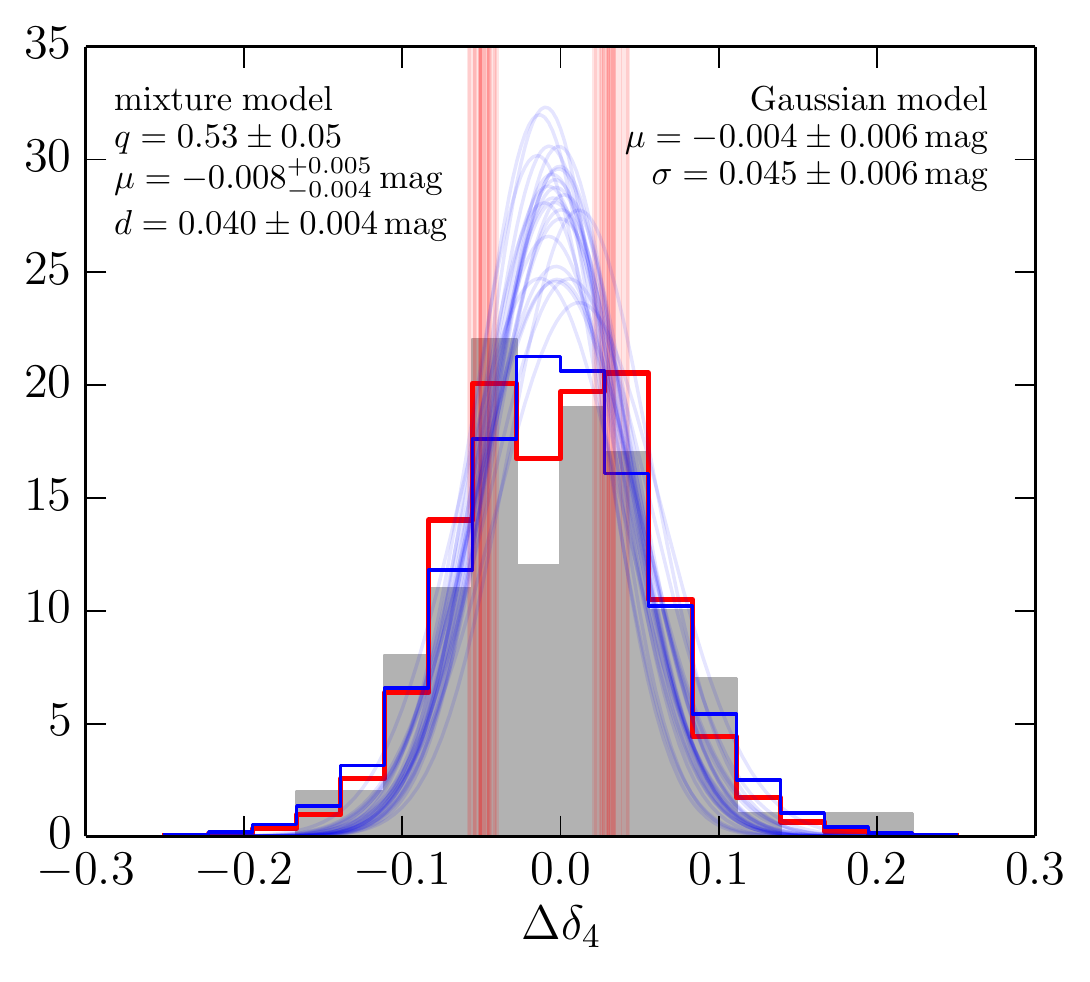}
\caption{Observed distribution of the ridge-line-detrended $\delta_4$ index for stars brighter than $y_0=19$\,mag (grey histogram) and two different models for the intrinsic distribution. Red vertical lines represent a sampling from the posterior pdf of the location of the two populations with no intrinsic spread in the mixture model (Eq.~\ref{eq:mixturemodel}). Blue transparent curves show a sampling from the posterior pdf of the intrinsic Gaussian distribution model. Best-fitting values for the parameters of each model in terms of the median of the posterior and 68\,per cent credible intervals are shown at the top. The solid red and blue lines correspond to the best respective models convolved with the photometric uncertainties and integrated over the bins of the histogram. A leave-one-out cross-validation to quantify which model fits better yields relative odds of $\sim\!$5.1 to one for the mixture and Gaussian models (see text).}
\label{fig:d4bimodal}
\end{figure}

The results for the two models, i.e. an intrinsic Gaussian distribution and the mixture of two components with no intrinsic spread, are shown in Fig.~\ref{fig:d4bimodal} where a histogram of the observed $\Delta\delta_4$ distribution is overlaid with samplings from the posterior pdf of both models for the \emph{intrinsic} distribution, i.e. before convolution with the photometric uncertainties. The mixture model suggests two populations of similar size  ($q=0.53\pm0.05$) separated by $d=0.041\pm0.004$\,mag. For comparison, previous studies of NGC~2419 favour a smaller fraction of second-generation stars of $q=0.3-0.4$ \citep{2012MNRAS.426.2889M,2013MNRAS.431.1995B,2013ApJ...778L..13L}. A possible explanation for this difference is in the different radial distributions between our sample and the literature samples. While it is difficult to quantify the selection function of the literature studies, it is very plausible that our sample in comparison is biased towards larger radii because our photometry avoids a comparatively large part of the cluster centre due to the mediocre seeing. In this case, however, finding a larger second-generation fraction $q$ in our data would suggest that the second generation in NGC~2419 is less centrally concentrated. We will come back to this possibility in Section~\ref{sec:disc}.

The solid red and blue curves in Fig.~\ref{fig:d4bimodal} correspond to the best-fitting models (in terms of the median of the posterior pdfs)  convolved with the photometric uncertainties and integrated over the same histogram bins as the observed data. Despite the fundamentally different intrinsic distributions, both error-convolved curves qualitatively reproduce the observations similarly well. In order to quantify which model fits better, we perform a leave-one-out cross-validation to determine the relative odds of both models following the prescription of \citet[][]{2014MNRAS.443..815F}. Briefly, the procedure determines the probability of observing a left-out data point given a model fit to the remaining data points; the product of the probabilities obtained by leaving out each data point in turn is a measure of the odds of a given model. Cross-validation guards against over-fitting, i.e. choosing an overly complex model, as a model with many degrees of freedom will tend to follow more closely the n-1 data points, but then will not perform as well in predicting the left-out data point. In the Bayesian formalism the individual leave-one-out probabilities can be evaluated directly from samples of the MCMC chain that was used to fit the model \citep{Gelfand1995}. For our data, this yields relative odds for the mixture and intrinsic Gaussian models of $\sim\!5.1$:1. Thus, the data slightly favour two distinct populations with no intrinsic spread over an intrinsic Gaussian distribution in $\Delta\delta_4$. However, the model odds are similar enough that -- based on our photometric data alone -- we are not inclined to \emph{claim} the existence of two distinct populations. 

\section{Summary and discussion}
\label{sec:disc}
We obtained wide-field \Sgren $uvby$ photometry of the outer halo GC NGC~2419. Classifying sources in our photometric catalogue based on their gravity- and metallicity-sensitive \Sgren $c_1$ and $m_1$ indices resulted in a catalogue of 419 candidate cluster members that is virtually free of contaminants (estimated contamination $\sim1$\,per cent).

The metallicity of sources classified as likely RGB stars was determined based on four different photometric metallicity calibrations from the literature. Using the subset of 114 clean RGB stars brighter than $y=19$\,mag and a realistic, non-Gaussian model for the photometric uncertainties, we inferred the cluster's intrinsic metallicity distribution function. The metallicity relations of \citet{2000A&A...355..994H}, \citet{AdenThesis}, and the semi-empirical calibrations by \citet{2007ApJ...670..400C} using $(v-y)_0$ and $(u-y)_0$ colours, yield a cluster mean [Fe/H] of $\mu_\mathrm{intr,H00}=-1.96\pm0.01$, $\mu_\mathrm{intr,A11}=-2.43^{+0.03}_{-0.02}$,  $\mu_\mathrm{intr,C07vy}=-2.07\pm0.02$, and $\mu_\mathrm{intr,C07uy}=-2.08\pm0.02$\,dex, respectively. The last two estimates are in excellent agreement with the cluster's mean spectroscopic metallicity of [Fe/H]=$-2.09\pm0.02$\,dex \citep{2012MNRAS.426.2889M,2012ApJ...760...86C}. For individual bright RGB stars, the \citet{2007ApJ...670..400C} relations also agree well with spectroscopic [Fe/H], while the other two calibrations show systematic offsets at the $\sim\!$0.2\,dex level.

For the cluster's intrinsic spread in [Fe/H] we measure ${\sigintr,}_\mathrm{H00}=0.11^{+0.02}_{-0.01}$, ${\sigintr,}_\mathrm{A11}=0.27^{+0.03}_{-0.02}$, ${\sigintr}_{\mathrm{C07}vy}=0.15^{+0.02}_{-0.01}$, and ${\sigintr,}_{\mathrm{C07}uy}=0.15\pm0.01$\,dex, respectively from 114 RGB stars brighter than $y_0=19\,$mag. Since the \citet{2000A&A...355..994H} iso-metallicity curves show the least systematic tilt compared to our data, we regard ${\sigintr,}_\mathrm{H00}=0.11^{+0.02}_{-0.01}$\,dex as the best formal estimate of NGC~2419's intrinsic metallicity spread. However, because the intrinsic precision of the photometric metallicity estimation and the effects of systematics are similar in magnitude, this is an upper limit to the cluster's true spread in [Fe/H]. The lack of a significant correlation between the photometric and spectroscopic [Fe/H] constitutes further evidence against a true intrinsic [Fe/H] spread on the $\ga$0.1\,dex level. Our results support the vanishing intrinsic spreads of $\sigma_\mathrm{[Fe/H]}=0.00\pm0.03$ and $0.00\pm0.08$\,dex inferred by \citet{2012MNRAS.426.2889M} based on direct spectroscopic measurements of [Fe/H] of 49 and 7 RGB stars in NGC~2419, respectively. 

The CN-sensitive \Sgren indices $\delta_4$ and $c_y$ (among others) of NGC~2419's RGB stars show strong correlations or anti-correlations with spectroscopic [Mg/Fe] and [K/Fe] abundance, suggesting that the Mg-deficient second generation is enhanced in nitrogen. The spreads in these CN-sensitive indices of $0.04-0.05$\,mag are comparable to those found in other  metal-poor Galactic GCs. Despite their sensitivity to CN, some indices and colours, such as $m_1$ and $(u-y)$, show no correlation with spectroscopic [Mg/Fe] and [K/Fe]. This supports previous findings of a substantial He enhancement in the second generation that, in these colours and indices, approximately compensates the offsets caused by nitrogen enhancement. 

We tested for the existence of two distinct populations in the $\delta_4$ distribution of bright RGB stars, because this index correlates most strongly with [Mg/Fe] and [K/Fe], and is strongly dependent on nitrogen abundance, but independent of He. We found that the data prefer a model comprised of two distinct populations with no internal spread in $\delta_4$, with the nitrogen-enhanced population making up roughly half of the stars $q=0.53\pm0.05$. However, based on a cross-validation estimate of the relative model odds, a continuous Gaussian distribution in $\delta_4$ with finite intrinsic dispersion was found to be only marginally less likely ($\sim\!$0.2:1).

The nitrogen-sensitivity of the \Sgren $u$-band at low metallicity is mainly due to the 3400\AA\ NH molecular that also falls within the SDSS-like $u$ band filter used by \citep{2013MNRAS.431.1995B}, so that it is possible that the spread in $u_\mathrm{SDSS}-V$ reported by these authors is in part also driven by nitrogen rather than He alone. The magnitude of both contributions will depend on the stellar luminosity, so that in principle, the net effect could change sign along the RGB. This may solve the riddle of why the Mg-deficient likely second generation stars close to the tip of the RGB are redder in $u_\mathrm{SDSS}-V$, while they should be bluer if the colour spread were due to He enhancement. On the other hand, if nitrogen is the primary source of the $u_\mathrm{SDSS}-V$ spread this may mean that the second generation in NGC~2419, unlike in most other clusters, is the radially more extended one, a possibility pointed out by \citet{2015ApJ...804...71L}, who found exactly this to be the case in the centre of M\,15. The somewhat higher fraction of second-generation stars of $q=0.53\pm0.05$ found in our data that are weighted towards larger radii (compared to q=0.3-0.4 reported in the literature based on more centrally weighted samples) would support this case. Detailed modelling of the relative effects of nitrogen and He enhancements would help to clarify this issue, as would a larger sample of high-quality spectra of NGC~2419 member stars. 

To this end, our catalogue of likely member stars presents an excellent target list for spectroscopic follow up, also for the purpose of constraining further the question of dark matter around NGC~2419 and its radial profile of velocity anisotropy \citep{2011ApJ...738..186I}. Thanks  to its remote location and therefore possibly undisturbed evolution, NGC~2419 is a promising target that can be used to address the general question of whether globular clusters are associated with dark matter halos. While recent studies of the cluster's internal dynamics have yielded no significant detection of dark matter \citep{2009MNRAS.396.2051B,2011ApJ...738..186I}, one of the limitations of current radial velocity samples is their concentration towards the cluster centre, where the stellar component dominates any possibly extant dark matter. Our catalogue lists $\sim$150 likely RGB stars brighter than $V\approx y<20$\,mag in the radial range $0.85<r<10.5$\,arcmin from the cluster centre, for which no radial velocity measurements exist.

While NGC~2419's remote location suggests that it may be an accreted cluster, its small or absent variation in [Fe/H] makes it different from other suspect accreted \emph{nuclei}, such as $\omega$Cen and M\,54. We note that it is not required that nuclear star cluster show a spread in [Fe/H], since it  may have formed, for example,  in a low-mass host galaxy that did not undergo significant chemical enrichment or was unable to form several globular clusters that could sink into  its centre via dynamical friction, and merge. However, it is unlikely that the relatively massive NGC~2419 was born in a low-mass quiescent dwarf galaxy. Its size and luminosity are similar to nuclear star clusters of dwarf elliptical galaxies as luminous as M$_V\!\sim\!-12$\,mag \citep[e.g.][ where we approximated a colour of $V-F814W\approx1$\,mag for both NGC~2419, and the nucleated dwarf ellipticals]{2014MNRAS.445.2385D}. This is comparable to the more massive early-type Milky Way satellites, such as the non-nucleated Sculptor or Leo I dwarfs \citep{2012AJ....144....4M}, both of which \emph{did} undergo significant chemical enrichment \citep[][]{2012A&A...539A.103D,2007ApJ...657..241K}. 

Conversely, it has also been suggested that massive (M$\ga10^6$\,M$_\odot$) GCs, even without being embedded as a nucleus in a dwarf galaxy, were able to retain the ejecta of type II supernovae and therefore were able to self-enrich in iron on the $\sim\!0.1$\,dex level \citep{2012AJ....144...76W}, although it is unclear whether this mechanism can account for extreme cases such as $\omega$Cen \citep{2014A&A...567A.105F}. The fraction of retained metals according to a simple self-enrichment model by \citet{2009ApJ...695.1082B} depends on the cluster's primordial radius as $\mathrm{exp}\left(-r\right)$. If we take the present-day sizes of GCs to be indicative of their relative sizes at birth, a reasonable assumption for GCs with M$\ga10^6$\,M$_\odot$ if the expansion is driven by internal evolution \citep{2010MNRAS.408L..16G}, we qualitatively expect a vanishing [Fe/H] spread in NGC~2419 compared to typical more compact (by a factor of $\sim\,5$) GCs of similar mass. Depending on the formation scenario of second-generation stars, similar considerations will apply to the retention fraction of light elements, suggesting that NGC~2419 -- thanks to its extended structure -- may turn out to be a unique test case for these scenarios.

In summary, NGC~2419 is unlikely to be a stripped nucleus, and apart from its large size and peculiar Mg, K, and potentially Ca abundance patterns that may be interrelated with its size, it appears to be a `normal' globular cluster, with a substantial second generation that is enhanced in nitrogen on a level similar to other metal-poor Galactic GCs.

\section*{Acknowledgements}
We thank Daniel Ad{\'e}n for his help with the observations and Frank Grundahl for his input at early stages of this project. M.J.F. would like to thank Hans-Günter Ludwig and Peter Stetson for fruitful discussions. M.J.F., A.K. and N.K gratefully acknowledge support from the DFG via Emmy Noether Grant Ko 4161/1. This research has made use of the \textsc{SIMBAD} database and the \textsc{VizieR} catalogue access tool, operated at CDS, Strasbourg, France and described in \citet{2000A&AS..143....9W} and \citet{2000A&AS..143...23O}, respectively. This research made use of \textsc{Astropy}, a community-developed core Python package for Astronomy \citep{2013A&A...558A..33A}.

\end{document}